\documentclass[12pt]{article}
\usepackage[utf8]{inputenc}
\usepackage[T1]{fontenc}
\usepackage{graphicx}
\usepackage{float}
\usepackage{array}
\usepackage{url}
\usepackage{hyperref}
\usepackage{authblk}
\usepackage{amsmath}

\title{Assessing the Reliability of Large Language Models for Deductive Qualitative Coding: A Comparative Study of ChatGPT Interventions}

\author[1]{Angjelin Hila\\\href{https://orcid.org/0009-0005-0481-0443}{ORCID: 0009-0005-0481-0443}}
\author[2]{Elliott Hauser\\\href{https://orcid.org/0000-0002-2547-0952}{ORCID: 0000-0002-2547-0952}}

\affil[1]{School of Information, University of Texas, Austin\\\texttt{ahila@utexas.edu}}
\affil[2]{School of Information, University of Texas, Austin\\\texttt{eah13@utexas.edu}}

\date{}  

\begin{document}  

\maketitle        

\sloppy
\begin{abstract}
In this study, we investigate the use of large language models (LLMs), specifically ChatGPT, for structured deductive qualitative coding. While most current research emphasizes inductive coding applications, we address the underexplored potential of LLMs to perform deductive classification tasks aligned with established human-coded schemes. Using the Comparative Agendas Project (CAP) Master Codebook, we classified U.S. Supreme Court case summaries into 21 major policy domains. We tested four intervention methods: zero-shot, few-shot, definition-based, and a novel Step-by-Step Task Decomposition strategy, across repeated samples. Performance was evaluated using standard classification metrics (accuracy, F1-score, Cohen’s $\kappa$, Krippendorff’s $\alpha$), and construct validity was assessed using chi-squared tests and Cramér’s V. Chi-squared and effect size analyses confirmed that intervention strategies significantly influenced classification behavior (e.g., V = 0.613 for definitions vs. few-shot). The Step-by-Step Task Decomposition strategy achieved the strongest reliability (accuracy = 0.775, $\kappa$ = 0.744, $\alpha$ = 0.746), achieving thresholds for substantial agreement. Despite the semantic ambiguity within case summaries, ChatGPT displayed stable agreement across samples, including high F1 scores in low-support subclasses. These findings demonstrate that with targeted, custom-tailored interventions LLMs can achieve reliability levels suitable for integration into rigorous qualitative coding workflows.
\end{abstract}
\fussy 

\maketitle

\section{Introduction}
The recent advent of LLM chatbots such as OpenAI's ChatGPT, Google's Gemini, and Anthropic's Claude, to name the most prominent and recognizable, have prompted interest in their integration within qualitative coding flows. The thrust of this interest rests on efficiency advantages that chatbots can provide in comparison to existing automated and semi-automated coding tools such as Nvivo, Atlas.ti and nCoder. In this paper, we contribute to the growing body of research on qualitative coding with chatbots. Specifically, we situate our contributions within the domain of structured deductive coding tasks, an area that remains underexplored by existing research. We utilize the Comparative Agendas Codebook (CAP) as a human-coded and validated instrument designed to classify international policy agendas with the aim of reproducing coding accuracy and reliability with ChatGPT. In doing so, we seek to ascertain whether ChatGPT and, by extension, comparable models can be considered as equivalent to human coders or as coding assistants that improve coding scalability and reliability. 

Qualitative coding constitutes a major methodological component of qualitative data analysis. In general, qualitative coding describes a collection of methodologies that aim to draw patterns from qualitative data in the form of codes or categorical schemes. These fall into two broad categories: inductive and deductive coding. Inductive coding methods derive codes directly from the data without predefined categories or theories. Deductive coding, by contrast, involves the application of predefined codes or categories, based on existing theories, frameworks, or hypotheses, to the data \cite{saldana2015}. Qualitative data can be highly heterogeneous, ranging from interviews, focus groups, observations, ethnographies, case studies, to graphical, text and online data. Because the aims of qualitative coding vary across studies and research objectives, the coding process can be highly interpretative. Distinct coding techniques include in vivo coding, process coding, open coding, descriptive coding, structural coding and value coding \cite{saldana2015}. Hybrid approaches combine both inductive and deductive coding steps whereby the inductive procedure generates a scheme that is subsequently applied to deductively classify new instances of data. Across coding techniques, the process is refined through iteration and involves at least two steps: initial coding and line-by-line coding \cite{saldana2015}. 

Two major challenges beset qualitative coding: a) scalability b) reliability.  Because coding tasks were traditionally performed by human coders, time constraints made scalability difficult, while the potential for human error and inter-coder variation presents challenges for the consistency and validity of codebooks. Scalability, therefore, refers to the problem of deriving codes from, and applying codes to, large datasets \cite{trilling2018, chung2020}. The problem of scalability affects both inductive and deductive coding procedures. Reliability, on the other hand, concerns the consistency of codes across human coders,  termed inter-coder or inter-rater reliability \cite{hallgren2012, chaturvedi2015, gwet2021}. In contrast to inter-rater agreement, which measures the raw percentage agreement between raters or coders, inter-rater reliability adjusts the absolute agreement between rater values to reflect consistency in coding patterns beyond chance \cite{chaturvedi2015}. High thresholds of inter-rater reliability between coders should not be confused with the construct validity of the coding scheme, although they indirectly contribute to construct validity by contributing to the reproducibility of the results \cite{hallgren2012, chaturvedi2015, gwet2021}. 

To mitigate these problems, automated coding techniques such as Computer-Assisted Qualitative Data Analysis Software (CAQDAS) were proposed in the 80s and 90s \cite{Bringer2006, sinkovics2012, olapane2021}. These methods facilitated systematic coding procedures rather than automating the entire process. During the 2000s and 2010s, automated coding procedures leveraged natural language processing (NLP) and machine learning (ML) techniques \cite{sinkovics2012, nelson2018}. These ranged between dictionary and rules-based methods to direct coding through artificial neural networks (ANNs) \cite{Bringer2006, sinkovics2012, nelson2018}. Dictionary and rule-based methods faced limitations such as rigidity, ambiguity, and poor context-awareness \cite{cypress2019, orr2024}. While ANN-based methods addressed some of these shortcomings by reducing manual-effort and increasing contextual understanding,  they remained heavily dependent on feature engineering, poor transferability, and handling of global context \cite{orr2024}. 

The emergence of the transformer architecture and large language models (LLMs) opened a gambit for improved automated coding by balancing scalability and context-awareness without requiring feature engineering \cite{schmidt2021, zambrano2023}. Since the introduction of ChatGPT, publicly available chatbots have been suggested as tools for both deductive and inductive coding tasks \cite{morgan2023, zambrano2023, bijker2024, wachinger2024}. A number of recent studies have demonstrated the utility of ChatGPT and LLM chatbots more broadly to aid human coders in both inductive and deductive coding tasks \cite{bijker2024, gao2024, gebreegziabher2023, zambrano2023, morgan2023, wachinger2024, vatsal2023}. However, the majority of studies focus primarily on inductive coding and less on deductive coding. Due to this focal asymmetry greater uncertainty persists with respect to optimizing the scalable integration of chatbots within deductive qualitative coding tasks that meet standard thresholds of reliability. 

Accordingly, in this paper we address the following research questions: \\
\vspace{2mm}
\textbf{Q1:} What is the baseline reliability of ChatGPT in deductive coding? \\
\vspace{1mm}
\textbf{Q2:} Are there distinct prompt engineering interventions and coding pipelines that improve baseline benchmarks? \\
\vspace{1mm}
\textbf{Q3:} Can an LLM chatbot serve as a qualitative coding assistant? \\

\section{Related Research}

Recent research has proposed incorporating LLMs into qualitative coding in order to reduce costs of manual reliability-enhancing measures such as triangulation, collaborative coding, member checking, and consensus-building \cite{bijker2024, gao2024, gebreegziabher2023, zambrano2023, morgan2023, wachinger2024}. First we review recent tools aimed at automating or semi automating qualitative coding. Second, we review recent work that aims to incorporate ChatGTP and comparable chatbots into qualitative coding tasks. 

The advent of LLMs has renewed interest in optimal methods of hybridizing human and AI intelligence in qualitative coding tasks. Rietz and Maedche \cite{rietz2021} developed Cody, an interactive user-facing QDA for semi-automated coding that mixes rule-based with supervised learning, in order to address interaction and transparency shortcomings in extant QDA tools such as MAXQDA, Nvivo, Atlas.ti, Dedoose, WebQDA, and QDAMiner where the system restricts user interactions to accepting or rejecting input. Rietz and Maedche find that a combination of manual and automated annotations achieves the most complete and accurate results \cite{rietz2021}. Cody enables users to specify their desired unit-of-analysis, add annotations and codes, define coding rules, react to suggestions, and access rudimentary statistics \cite{rietz2021}. More recently, Gebreegziabher et al. \cite{gebreegziabher2023} developed PATAT, an AI-enabled tool that uses an interactive program synthesis approach to learn flexible and expressive patterns over user-annotated codes in real-time as users annotate data. In addition, Gao et al \cite{gao2024} designed CollabCoder, an AI tool that generates code suggestions during the independent open coding phase in order to promote consensus-building for the iterative discussion phase and codify disagreements between researchers. While these tools constitute progress in automating various aspects of qualitative coding, they lack the dialogical flexibility and context sensitivity that AI chatbots enable through interactive, interpretive exchanges. 

Concurrent to custom-made tools, a growing dimension of research focus consists of enlisting LLM chatbots into qualitative coding tasks. Zambrano et al \cite{zambrano2023} compared ChatGPT performance to nCoder across 2 binary sentiment codes on press release data. They found nCoder outperforms ChatGPT average Kappa agreement with humans in precision, but that ChatGPT outperformed nCoder in recall \cite{zambrano2023}.  More generally, they found that ChatGPT offers high performance in code categories where the range of possible interactions is limited and the semantic field is concrete and conclude that ChatGPT can add explainability in inductive coding that can help improve human coder consistency and construct validity. Morgan \cite{morgan2023} sought to determine whether ChatGPT would reproduce themes derived from human conducted reflexive thematic analysis. He found that ChatGPT performed well at descriptive and concrete themes but was less successful at locating subtle or interpretive themes.  Wachinger et al \cite{wachinger2024} compared human and ChatGPT performance in qualitative coding interviews based on grounded theory, reflexive thematic analysis and five step framework approach, finding that ChatGPT displayed considerable overlap with human analysis, and was able to suggest codebooks with face validity, and justify its coding choices with reference to specific theory. 

These promising findings are further buttressed by quantitative analysis of LLM chatbot and human coder agreement. Bijker et al. \cite{bijker2024} more recently tested the reliability of ChatGPT for inductive and deductive content analysis coding tasks. They tested three coding techniques in total, specifically data-driven inductive coding, unconstrained deductive coding, and deductive coding using the Theoretical Domains Framework (TDF) on a dataset comprising online posts sourced from various online forums, social media and professional platforms on the topic of experiences reflecting changes on sugar consumption \cite{bijker2024}. ChatGPT was instructed to conduct each coding task 10 times. They found that precision rates between ChatGPT and human coding ranged between 0.66-0.88 \cite{bijker2024}. They further found that, across ChatGPT iterations per coding task, the $K$\text{-score} score ranged between 0.58 and 0.95. ChatGPT performed best at the inductive coding task yielding an average $K$\text{-score} of 0.84, second best at the unconstrained deductive task yielding a $K$\text{-score} of 0.73 and third best at the structured deductive task yielding an average $K$\text{-score} of 0.66 \cite{bijker2024}. The study did not however, explicitly compare human and ChatGPT $\text{K-score}$, underscoring a major limitation of the study. Further, even though the study applied an iterative prompt engineering process to zero-in on prompts that yielded desired results, the refinement process was informally conducted. 

A further dimension in the automation of qualitative coding involves understanding the situated needs and workflows of professional researchers.   Seeking to determine whether first-pass qualitative coding could be partially automated, Marathe and Toyama \cite{marathe2018} interviewed 15 qualitative researchers across two rounds in an academic setting and found that researchers follow common practices such as consistent use of units of analysis, multilevel code construction, and iterative codebook development suitable to automation, but prefer automation only after developing the codebook and coding a subset of the data. These findings support our concern with the reliable and scalable application of preexisting structured deductive coding schemes to large bodies of qualitative data. 

Orthogonal to research on chatbot-assisted and automated coding, a growing body of research recognizes the need for reporting interrater reliability (IRR) metrics. Mcdonald et al \cite{mcdonald2019} find that IRR reporting occurrs in only 1/9 of qualitative papers. In this study, we utilize Cohen’s Kappa \cite{cohen1960} and Krippendorff’s Alpha \cite{krippendorff2011} as measures of intercoder reliability. Following Landis and Koch and McHugh \cite{landis1977, mchugh2012}, we interpret Kappa values of 0.61–0.80 as indicating substantial agreement and values above 0.80 as almost perfect agreement. For Krippendorff’s Alpha, we adopt the thresholds proposed by Krippendorff \cite{krippendorff2011}, treating values above 0.800 as reliable, values between 0.667 and 0.800 as indicating substantial agreement for tentative conclusions, and values below 0.667 as insufficient for drawing reliable inferences. These thresholds serve as the interpretive benchmarks for assessing the consistency and reliability of coding across intervention methods in our study.

\section{Data}

In line with our goal of integrating chatbots into structured, deductive coding tasks, we adopted the \textit{Comparative Agendas Project Master Codebook} (CAP MC): a human-curated, hierarchically structured codebook validated for cross-national, cross-cultural policy analysis \cite{baumgartner2019, walgrave2019}. The Comparative Agendas Project (CAP) originated from the U.S. Policy Agendas Project (US PAP), developed by Frank Baumgartner and Bryan Jones, which aimed to systematically trace the policy content of governmental and public agendas over time \cite{walgrave2019}. CAP has since evolved into a global network of country-specific agenda projects, each using a shared classification system to enable comparative policy research across contexts.

The CAP Master Codebook was developed and coordinated by Shaun Bevan, who directed the Master Codebook Project at the University of Edinburgh \cite{walgrave2019, bevan2020}. Bevan collaborated with project leads from national teams to harmonize locally generated codebooks into a unified coding scheme through an iterative, comparative method known as \textit{crosswalking} \cite{dowding2016, bevan2020}. This process involves aligning and mapping bottom-up, country-specific policy codes into a common framework of major and subtopic categories \cite{dowding2016}. Research into cross-national qualitative coding shows that transnational coding clusters can significantly enhance interpretive validity by addressing linguistic nuance, power dynamics, and contextual meanings in multilingual data \cite{rodriguez2022}. 

The CAP MC comprises 21 major policy topics, each of which decomposes into up to 10--15 subtopics, totaling approximately 220 subcategories. It has been widely adopted in comparative policy research due to its conceptual clarity, replicability, and demonstrated inter-coder reliability \cite{dowding2016}. Given these strengths, we leveraged the CAP MC as a coding standard for developing and evaluating chatbot-assisted deductive coding. Specifically, its standardized, validated structure serves two key purposes: (a) it provides human-validated data for measuring human-chatbot agreement,(b) a ground-truth reference for evaluating chatbot classification performance, and (c) it offers access to a diverse range of human-coded datasets for testing the reliability and generalizability of automated classification methods.

Specifically, we chose the Supreme Court Cases dataset available from the (CAP) webapge {\url{https://www.comparativeagendas.net/datasets_codebooks}} , which is the only publicly available dataset to examine the Courts agenda from a policy perspective \cite{cap_supreme_court}
. The dataset codes each case by its policy content and includes additional variables such as the Court’s ruling, where applicable \cite{cap_supreme_court}
. The dataset contains 10236 observations spanning the years 1901 to 2023. For our study, we only focused on three variables: (a) Supreme Court Case summaries, (b) CAP major topic label, and (c) CAP subtopic label \cite{cap_supreme_court}
. 

\subsection{Data Preprocessing}

The original datasest has a shape of 10236 rows and 17 attributes (10236, 17). After removing missing values, duplicates, and summaries that did not meet the context threshold of at least two sentences, the dataset was pruned to 9330 tuples and 3 relevant attributes (9330, 3). The attributes of interest were the Supreme Court Case summaries, and the CAP labels. 

The CAP codebook is a hierarchical taxonomy with two levels: a) major topic of 21 classes and b) subtopic of 220 classes. Each major topic contains approximately 10 subtopics. 

Because the dataset is complete with respect to major topic but incomplete with respect to subtopic (i.e. it has missing values), we removed incomplete data points because the ground-truth label is unknown for those instances. We also omitted summaries that did not meet a length threshold of two sentences appropriate for coding. Further, because the original dataset labels were integers, we used a mapping function to generate corresponding categorical labels. We matched each numerical label major topic and subtopic with its corresponding categorical label. 

The rationale for using categorical labels is that ChatGPT performs much more poorly with numerical labels. We found that ChatGPT performs significantly better with categorical labels than numerical labels. We believe this is the case because ChatGPT categorical contextual understanding supersedes its ability to learn numerical associations with text data. As a result, ChatGPT can leverage its prior contextual semantic understanding to apply nominal labels, whereas for numerical labels it has to learn the associations from scratch. After assigning a major label and sublabel to each data point, we randomized the dataset with the md5 hash function. 

\section{Methods} 
\begin{figure}[H]
    \centering
    \includegraphics[width=0.6\textwidth, keepaspectratio]{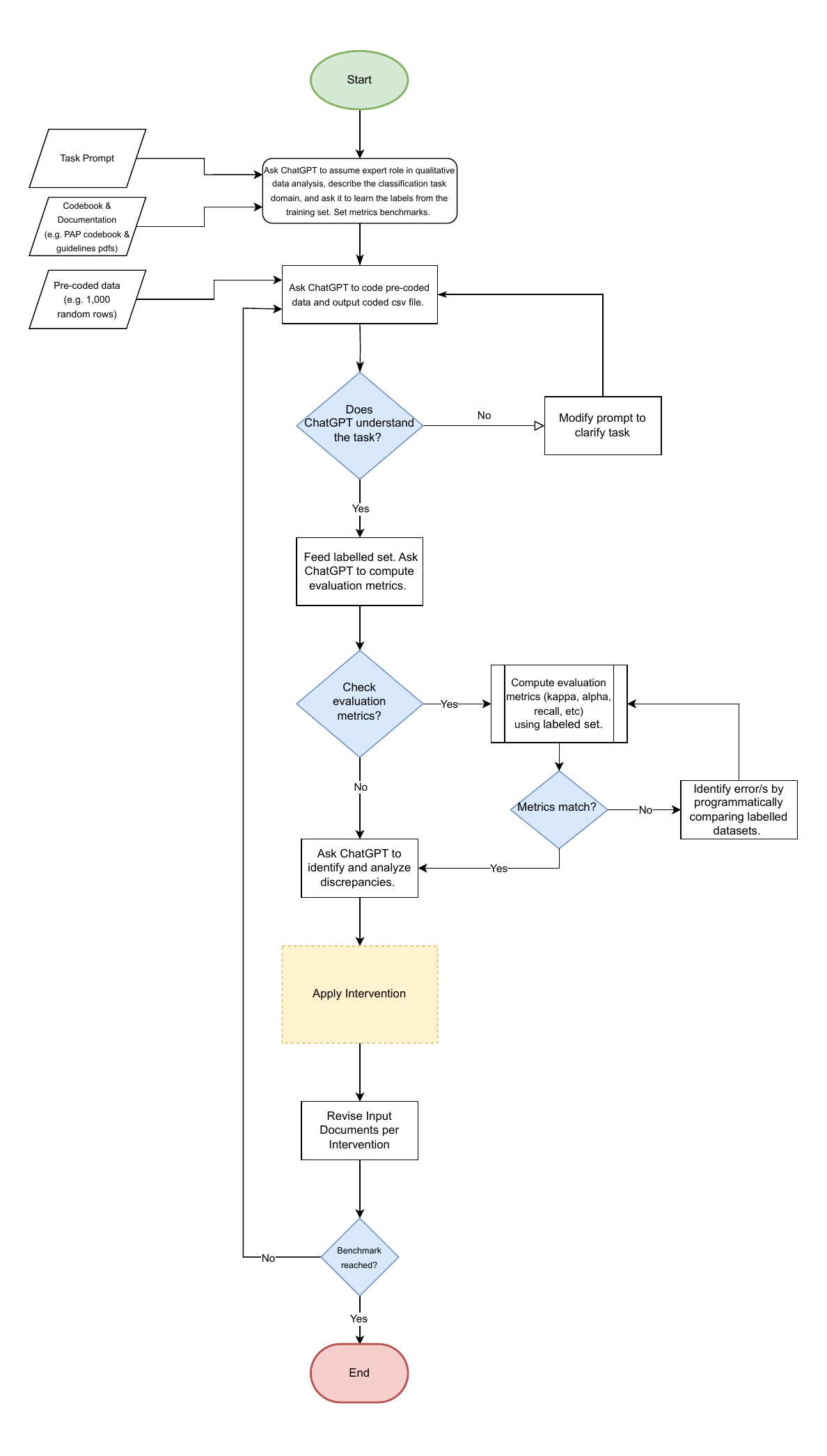}
    \caption{ChatGPT Coding Task Flowchart}
    \label{fig:Pipeline}
\end{figure}

We began by establishing baseline classification performance for ChatGPT using a pre-labeled dataset of U.S. Supreme Court case summaries annotated with major policy domains from the Comparative Agendas Project (CAP) Master Codebook. The initial classification task was executed without interventions to set benchmark performance metrics, including accuracy, F1-score, and Cohen’s $\kappa$. Following baseline evaluation, we developed four intervention strategies designed to scaffold ChatGPT's reasoning process: (1) zero-shot prompting, (2) few-shot prompting, (3) definition-based prompting, and (4) a Step-by-Step Task Decomposition method applied on a case-by-case basis. Each method was applied across 30 stratified random samples of 50 cases each to ensure balanced representation of policy classes and to satisfy chi-squared assumptions. For each intervention, we computed interrater agreement and classification performance metrics relative to human-coded labels. We also conducted chi-squared tests to assess convergent validity (within-method agreement) and discriminant validity (between-method differences), and calculated Cramér’s V to estimate effect sizes. The Step-by-Step Task Decomposition method, in particular, was motivated by the need to align ChatGPT’s reasoning with human deductive logic by decomposing each classification into explicitly articulated steps. This pipeline, depicted in Figure~\ref{fig:Pipeline}, was iteratively refined through prompt engineering, metric validation, and error analysis until the intervention achieved substantial agreement and met reliability thresholds.

\subsection{Intervention Definitions}
To assess the reliability and accuracy of ChatGPT as a top-down qualitative coder, we implemented and validated four distinct intervention strategies, each designed to scaffold classification performance to varying degrees:

\begin{itemize}
    \item \textbf{Intervention 1:} \texttt{Zero-Shot.} This baseline approach provides ChatGPT with only the list of class names, without any accompanying definitions or examples. It relies solely on the model’s general semantic understanding to assign labels, reflecting an unassisted, minimal-prompting setup.

    \item \textbf{Intervention 2:} \texttt{Few Shot.} This method supplies the model with several fully labeled examples from the training dataset. By observing real case-label pairs, the model learns to generalize labeling patterns and apply them to new, unseen instances. This simulates basic inductive learning from exemplar data.

    \item \textbf{Intervention 3:} \texttt{Definition.} In this approach, ChatGPT is given formal class definitions and associated key indicators. The definitions clarify the semantic scope of each class, while keyword cues help constrain ambiguity and guide alignment between case summaries and categories.

    \item \textbf{Intervention 4:} \texttt{Interactive.} This strategy prompts ChatGPT to perform deductive classification by explicitly reasoning through each case. The model is instructed to link relevant textual evidence to specific class criteria and articulate its rationale before providing a final label. This case-by-case method most closely emulates a human coder’s interpretive process.
\end{itemize}

\subsection{Intervention Validation}
\subsubsection{Sampling Method}
To ensure the validity of our repeated-measures design, we employed stratified random sampling. The chi-squared ($\chi^2$) test assumes that at least 80\% of the expected frequencies in a contingency table are greater than or equal to 5. This assumption requires that each categorical value be sufficiently represented across samples to ensure statistical validity.

Given the skewed distribution of the class labels, we used stratified sampling to preserve the proportional representation of each class from the original dataset. This method prevents rare categories from being overrepresented or omitted.

Sampling was conducted in two stages. First, each major label was sampled proportionally. The stratified sample was adjusted to a fixed total size to ensure uniform sample sizes across trials, preventing disproportionate class representations. To determine the optimal sample size and number of samples, we performed a nested search: for each candidate value of $N$, beginning at $N = 500$ and progressively decreasing, we incrementally increased $n$ from zero until the chi-squared ($\chi^2$) assumption was satisfied.  This process identified \textit{n}=50 (sample size) and \textit{N}=30 (sample count) as the minimal values satisfying the assumption. 

Because we employed a repeated-measures, within-subjects design, stratified random sampling was sufficient to ensure equivalence across samples, circumventing the need for random assignment. We excluded classes with fewer than five observations for both major labels and sublabels. Given that our sampling procedure satisfies the $\chi^2$ assumptions and maintains the independence of observations, our dataset is suitable for construct validity testing and hypothesis testing across distinct intervention methods.

\subsection{Intervention Development and Optimization}

\subsubsection{Prompt Engineering}
First, we randomized the dataset and partitioned it into 10 parts of 1{,}000 tuples. We assigned a role to ChatGPT as an expert qualitative coder with a legal background and interactively apprised it of the classification class context. We provided ChatGPT with one of the randomized partitions as a training set. We then asked ChatGPT to code the subsequent nine datasets using only the case summaries. After generating a classified dataset, ChatGPT was provided with the ground-truth set and asked to compute evaluation metrics. These were also independently verified using standard Python-based evaluation tools to ensure accuracy. Metrics were generated on a per-class basis. ChatGPT was then queried on areas of low performance and instructed to analyze classes it frequently misclassified. After ChatGPT identified and explained discrepancies in its predictions, it was asked to generate a document of coding instructions that distilled performance insights and formulated generalizable decision heuristics from the feedback. This procedure was iterated across the full dataset until one complete classification epoch had been completed.

Following this initial phase, we applied a series of increasingly structured prompt engineering strategies. These began with minimalist zero-shot inputs and evolved through few-shot prompting and definition-based scaffolding. Ultimately, we developed a Step-by-Step Task Decomposition strategy that explicitly guided ChatGPT to reason through each case classification by referencing textual evidence and reflecting on decision criteria. These interventions were tested in a repeated-measures design and evaluated using classification metrics and interrater reliability measures. 

We found that performance improved marginally across iterations during the feedback loop, and substantially with the application of structured interventions. While category support often correlated with precision and F1-score, this was not uniformly true. To better understand performance variation, we examined the relationship between support and classification accuracy. When controlling for support, we observed that the semantic clarity and syntactic quality of the case summary significantly affected ChatGPT’s ability to classify accurately. This suggests that text quality, not just class frequency, plays an important role in chatbot-based deductive classification.

\begin{figure}[H]
    \centering
    \includegraphics[width=0.6\textwidth]{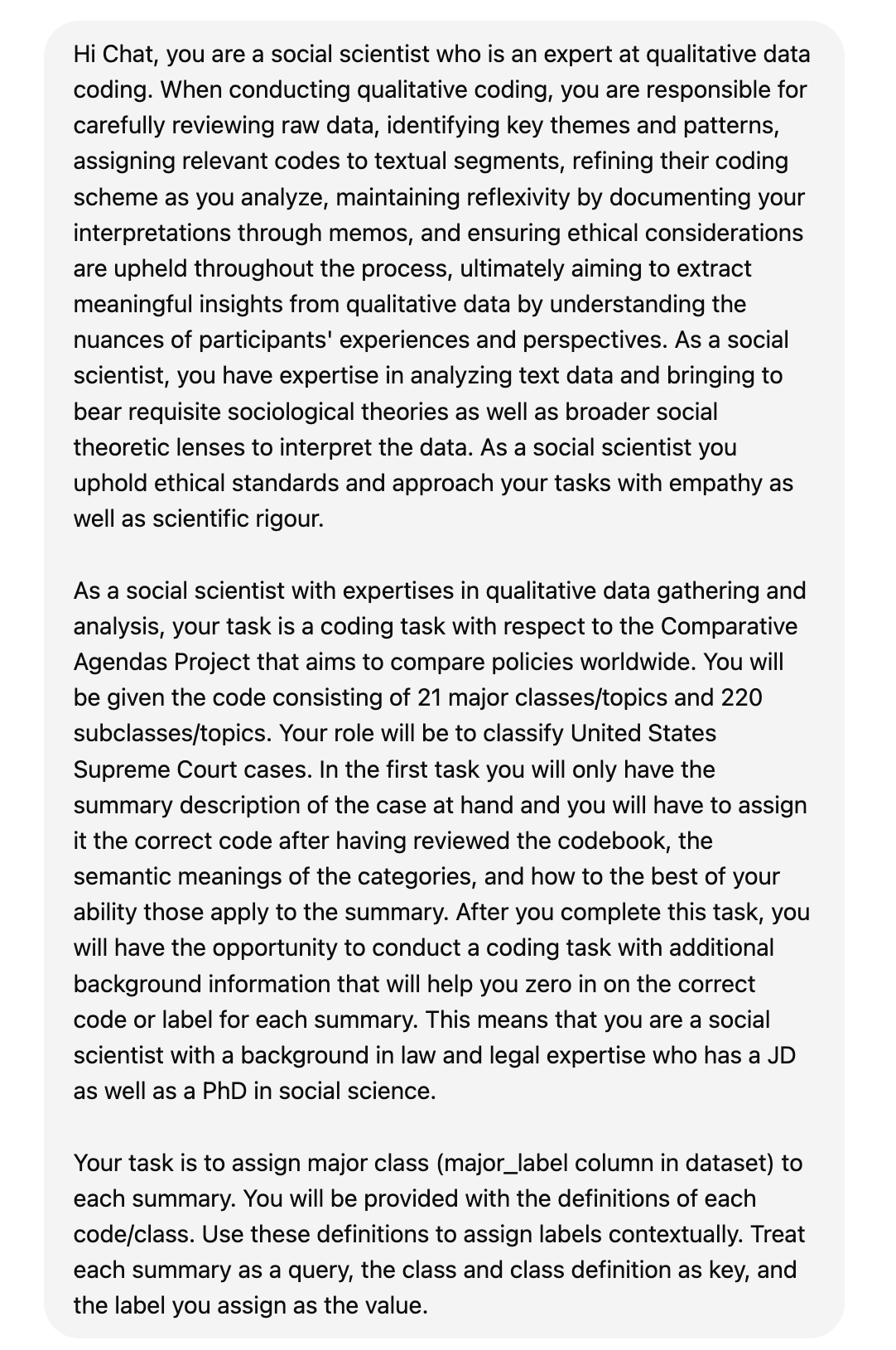}
    \caption{Step-by-Step Decomposition First Prompt.}
    \label{fig:ChatGPT_Prompt1}
\end{figure}

\begin{figure}[H]
    \centering
    \includegraphics[width=0.6\textwidth]{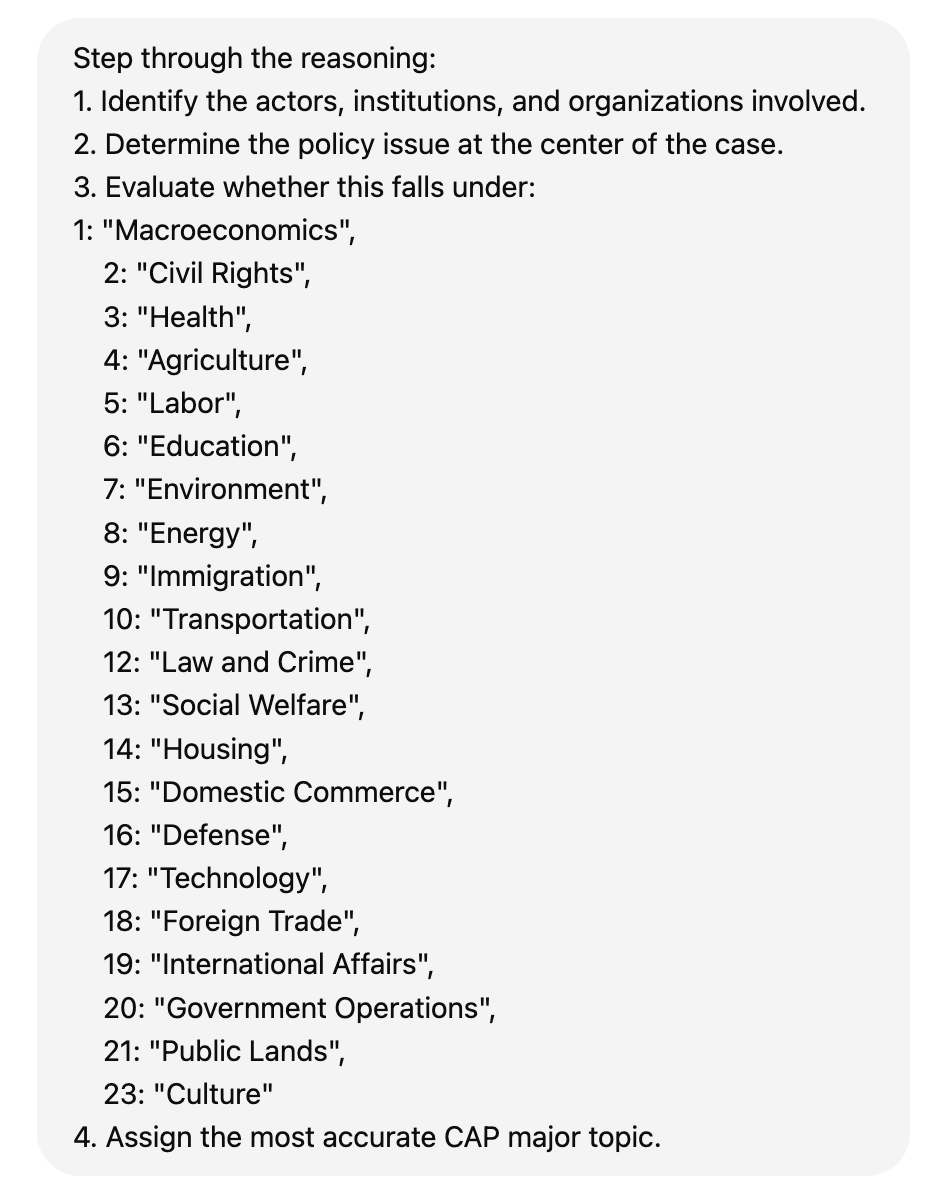}
    \caption{Step-by-Step Decomposition Second Prompt.}
    \label{fig:ChatGPT_Prompt2}
\end{figure}

Figure ~\ref{fig:ChatGPT_Prompt1} and ~\ref{fig:ChatGPT_Prompt2} show our prompts for Step-by-Step Task Decomposition. 

\subsubsection{Intervention Pipeline}

\subsection{Intervention Methods}
In order to formalize ChatGPT’s coding performance, we implemented four types of intervention strategies, each used to guide the model in classifying policy case summaries, aimed at improving performance and intercoder reliability. ChatGPT completed the coding task independently under each intervention condition. Below we describe each of the intervention methods. For this portion of our experiment, we focused exclusively on major class labels due to time and computational constraints, as well as the need to establish initial reliability benchmarks before extending the analysis to more granular subclass classifications. 

\subsubsection{\texttt{Zero-Shot.}}
For zero-shot classification, ChatGPT was assigned a role as a social scientist with expertise in qualitative coding and analysis, general knowledge about the CAP project and codebook, and instructions about the classification task. In this exercise, ChatGPT was provided with the major classes list without explicit definitions of each class. 

\subsubsection{\texttt{Few-Shot.}}
For few-shot classification, in addition to being assigned a role as a social scientist and apprised of the CAP project overview, ChatGPT was also provided with a training set. The training set consisted of two files comprising of 50 summaries along with the human-generated major class assignments. In this exercise, ChatGPT was provided with the major classes list without explicit definitions of each class. 

\subsubsection{\texttt{Definitions.}}
For  defintions classification, in addition to being assigned a role as a social scientist and apprised of the CAP project overview, ChatGPT was provided with the list of major classes and their concise definitions. In this intervention, ChatGPT was not provided with a training set, as we aimed to control definitions variable. 

\subsubsection{\texttt{Step-by-Step Decomposition}}
Finally, for step-by-step decomposition, ChatGPT was assigned a role as a social scientist, apprised of the CAP project overview, provided with a list of major classes, and told to step into the reasoning. Before classifying the test sets, ChatGPT was provided with samples of individual summaries and asked to identify the actors, institutions, and organizations involved, determine the policy issue at the center of the case, and evaluate whether it falls within one of the codified CAP major classes. Following these instructions, ChatGPT was cross-examined for three training instances. During each instance, ChatGPT was asked to reflect on the rationale for the label, support the rationale with evidence from the summary, and reassign the label if incorrect. For incorrect labels, ChatGPT's reasoning was compared to the textual evidence, and contradictory evidence from the text was provided. Finally, ChatGPT was asked to draw rules of thumb from these instances in order to classify future instances. 

\section{Results}

\subsection{Baseline Performance \& Reliability Benchmarking}

To evaluate the validity and consistency of automated classification, we rely on established benchmarks for inter-rater reliability. Cohen’s Kappa ($\kappa$) and Krippendorff’s Alpha ($\alpha$) are widely used measures for assessing agreement between coders beyond chance. According to conventional thresholds (Landis \& Koch, 1977), values between $0.61$ and $0.80$ represent \textit{substantial agreement}, while values above $0.80$ indicate \textit{almost perfect agreement}. Spearman’s rank correlation ($\rho$), though less frequently used for categorical agreement, offers a complementary view of ordinal association. For an automated coding system to be considered a reliable substitute for human coders, it should ideally reach or exceed these thresholds.

To establish such a benchmark, we compared ChatGPT’s out-of-the-box deductive coding performance to two supervised learning models, a bidirectional LSTM and RoBERTa, trained on the same dataset. As shown in Table~\ref{tab:model_performance}, ChatGPT achieved moderate performance on major labels (accuracy $= 0.57$, weighted F1 $= 0.52$, $\kappa = 0.46$), while its sublabel performance declined further ($\kappa = 0.41$). The LSTM performed comparably (major $\kappa = 0.55$), offering only marginal improvement. In contrast, RoBERTa achieved substantially higher reliability, with $\kappa = 0.75$ for major labels and $\kappa = 0.63$ for sublabels, placing it within the \textit{substantial agreement} range and thus serving as a practical upper-bound benchmark for automated performance. Notably, even RoBERTa's sublabel classification remained below the ideal reliability threshold.

These findings suggest that, without well-calibrated guidance, ChatGPT does not reach accepted standards of reliability and trails behind specialized models such as RoBERTa. However, its generative flexibility and reasoning capabilities enrich qualitative coding with explainability. Motivated by the gap between ChatGPT’s raw performance and benchmark expectations, we compared a series of structured intervention methods intended to improve the consistency and accuracy of its deductive reasoning. In particular, we sought to determine whether step-by-step feedback interventions elevate ChatGPT’s consistency and agreement with human-coded data to levels comparable to or exceeding supervised learning systems. By structuring the coding process around contextual cues, class definitions, and step-by-step reasoning, we hypothesize that ChatGPT can meet and potentially surpass conventional machine learning approaches in classification reliability.

\begin{table}[h]
    \centering
    \begin{tabular}{lccc}
        \hline
        \textbf{Model} & \textbf{Accuracy} & \textbf{Weighted F1} & \textbf{Cohen’s Kappa} \\
        \hline
        ChatGPT & 0.57 & 0.52 & 0.46 \\
        Bidirectional LSTM & 0.592 & 0.58 & 0.55 \\
        RoBERTa & 0.79 & 0.79 & 0.75 \\
        \hline
    \end{tabular}
    \caption{Major Label Model Performance Metrics}
    \label{tab:model_performance}
\end{table}

\begin{table}[H]
    \centering
    \begin{tabular}{lccc}
        \hline
        \textbf{Model} & \textbf{Accuracy} & \textbf{Weighted F1} & \textbf{Cohen’s Kappa} \\
        \hline
        ChatGPT & 0.46 & 0.4 & 0.41 \\
        Bidirectional LSTM & 0.21 & 0.58 & 0.244 \\
        RoBERTa & 0.65 & 0.63 & 0.63 \\
        \hline
    \end{tabular}
    \caption{Sub Label Model Performance Metrics}
    \label{tab:model_performance}
\end{table}

\subsection{Construct Validity: Convergent and Discriminant Analyses}
To assess the \textit{convergent} and \textit{discriminant validity} of our interventions, we conducted chi-squared ($\chi^2$) tests both within and between methods. Convergent validity was evaluated by calculating $\chi^2$ across samples produced by the same method, while discriminant validity was assessed by comparing samples across different methods.

The results show that within-method samples generally produced nonsignificant $\chi^2$ values ($p > 0.05$), indicating convergence, as shown in Table~\ref{tab:convergent_validity_trimmed}. In contrast, between-method comparisons yielded statistically significant $\chi^2$ results ($p < 0.05$), demonstrating divergence across interventions, as reported in Table~\ref{tab:discriminant_validity_trimmed}.

\begin{table}[h]
    \centering
    \resizebox{\textwidth}{!}{%
    \begin{tabular}{lcccccccc}
        \hline
        \textbf{Method} & \textbf{Mean Chi2} & \textbf{Std Chi2} & \textbf{Mean p value} & \textbf{Std p value} & \textbf{Significant p<0.05} & \textbf{Significant Bonferroni} & \textbf{Significant FDR} & \textbf{Total Tests} \\
        \hline
        Zero-shot & 149.480 & 36.260 & 0.482 & 0.340 & 70 & 3 & 20 & 435 \\
        Few-shot & 108.490 & 30.580 & 0.492 & 0.358 & 70 & 6 & 16 & 435 \\
        Definitions & 73.470 & 25.690 & 0.528 & 0.363 & 74 & 9 & 31 & 435 \\
        Interactive & 278.480 & 57.300 & 0.463 & 0.373 & 88 & 9 & 25 & 435 \\
        \hline
    \end{tabular}%
    }
    \caption{Convergent Validity: Statistical Summary for Within-Method Comparisons}
    \label{tab:convergent_validity_trimmed}
\end{table}

\begin{table}[H]
    \centering
    \resizebox{\textwidth}{!}{%
    \begin{tabular}{lcccccccc}
        \hline
        \textbf{Method} & \textbf{Mean Chi2} & \textbf{Std Chi2} & \textbf{Mean p value} & \textbf{Std p value} & \textbf{Significant p<0.05} & \textbf{Significant Bonferroni} & \textbf{Significant FDR} & \textbf{Total Tests} \\
        \hline
        Zero-shot vs Few-shot & 240.410 & 55.280 & 0.002 & 0.008 & 30 & 27 & 30 & 30 \\
        Zero-shot vs Definitions & 272.770 & 56.570 & 0.000 & 0.000 & 30 & 30 & 30 & 30 \\
        Zero-shot vs Interactive & 314.270 & 52.830 & 0.000 & 0.001 & 30 & 29 & 30 & 30 \\
        Few-shot vs Definitions & 148.650 & 48.760 & 0.050 & 0.117 & 23 & 18 & 23 & 30 \\
        Few-shot vs Interactive & 243.630 & 45.900 & 0.053 & 0.179 & 24 & 21 & 24 & 30 \\
        Definitions vs Interactive & 226.950 & 39.960 & 0.011 & 0.039 & 28 & 25 & 28 & 30 \\
        \hline
    \end{tabular}%
    }
    \caption{Discriminant Validity: Statistical Summary for Between-Method Comparisons}
    \label{tab:discriminant_validity_trimmed}
\end{table}

\subsection{Statistical Evaluation of Classification Interventions}
We used $\chi^2$ tests to evaluate whether the intervention methods produced significantly different distributions of major policy labels. All pairwise comparisons between methods yielded statistically significant results ($p < 0.001$), indicating that the classification outputs varied meaningfully across interventions. To assess the magnitude of these differences, we computed Cramér’s V for each method pair. The results suggest moderate to strong divergence in classification behavior. The strongest divergence was observed between the \textit{few-shot} and \textit{definition-based} methods ($\chi^2 = 1147.72$, $V = 0.613$), followed by the \textit{zero-shot} vs. \textit{few-shot} comparison ($\chi^2 = 779.08$, $V = 0.505$) and the \textit{definition-based} vs. \textit{interactive} pairing ($\chi^2 = 723.67$, $V = 0.487$). More moderate, though still substantial, differences were observed between \textit{few-shot} and \textit{interactive} ($V = 0.369$), and both \textit{zero-shot} vs. \textit{definitions} and \textit{zero-shot} vs. \textit{interactive} (each $V = 0.359$). These findings confirm that the intervention strategy employed has a significant and measurable effect on classification behavior, with effect sizes exceeding conventional thresholds for meaningful practical differences in categorical data analysis.

\vspace{2mm}

\begin{table}[H]
    \centering
    \begin{tabular}{l l r c r}
        \hline
        \textbf{Method 1} & \textbf{Method 2} & \textbf{Chi²} & \textbf{p-value} & \textbf{Cramér's V} \\
        \hline
        Zero-shot    & Few-shot      & 779.078   & $<$0.001 & 0.505 \\
        Zero-shot    & Definitions   & 392.335   & $<$0.001 & 0.359 \\
        Zero-shot    & Interactive   & 392.772   & $<$0.001 & 0.359 \\
        Few-shot     & Definitions   & 1147.722  & $<$0.001 & 0.613 \\
        Few-shot     & Interactive   & 415.845   & $<$0.001 & 0.369 \\
        Definitions  & Interactive   & 723.666   & $<$0.001 & 0.487 \\
        \hline
    \end{tabular}
    \caption{Pairwise chi-squared test results comparing intervention methods. All comparisons are statistically significant ($p < 0.001$), with Cramér's V indicating moderate to strong disagreement.}
    \label{tab:pairwise_chi_squared}
\end{table}

\paragraph{Q: Do Intervention Methods Differ Significantly from One Another?}
Cramér’s V values ranged from moderate ($V \approx 0.36$ for \textit{zero} vs. \textit{definitions} and \textit{interactive}) to strong ($V \approx 0.49$ for \textit{definitions} vs. \textit{interactive}) and very strong ($V \approx 0.61$ for \textit{few} vs. \textit{definitions}). These results provide robust evidence that intervention design substantially alters classification behavior, supporting the hypothesis that prompt-based methods diverge in their label assignments at both a statistically and practically meaningful level.

\begin{figure}[h]
    \centering
    \includegraphics[width=0.85\textwidth]{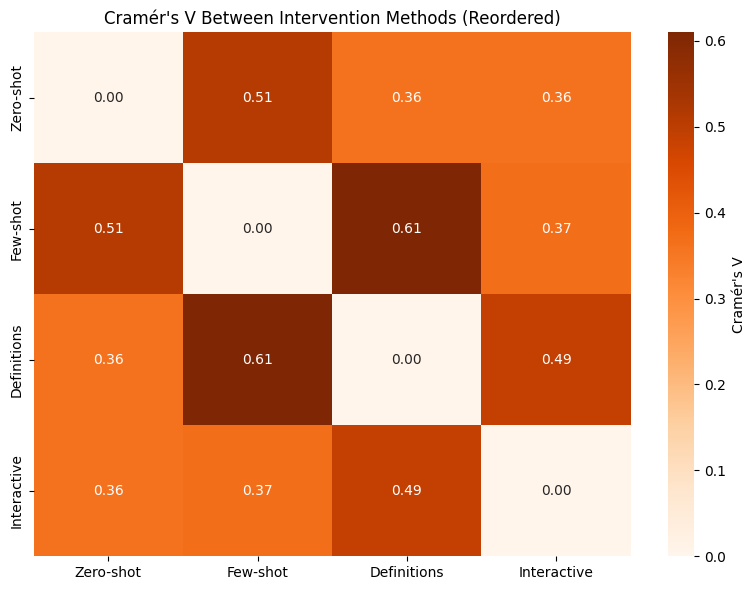}
    \caption{Cramér's V heatmap showing pairwise disagreement between intervention methods. Higher values indicate greater divergence in classification distributions.}
    \label{fig:cramer_heatmap}
\end{figure}

\vspace{2mm}
\paragraph{Q: Do intervention methods classify cases differently across major labels?} 

\begin{table}[htbp]
    \centering
    \small
    \begin{tabular}{lccc}
        \hline
        \textbf{Class} & \textbf{Chi²} & \textbf{p-value} & \textbf{Cramér's V} \\
        \hline
        \multicolumn{4}{l}{\textbf{High Disagreement} ($V \geq 0.20$)} \\
        \hline
        Government Operations & 807.627 & $< 0.001$ & 0.367 \\
        Law and Crime          & 437.382 & $< 0.001$ & 0.270 \\
        \hline
        \multicolumn{4}{l}{\textbf{Moderate Disagreement} ($0.10 \leq V < 0.20$)} \\
        \hline
        Domestic Commerce & 217.012 & $< 0.001$ & 0.190 \\
        Macroeconomics    & 213.691 & $< 0.001$ & 0.189 \\
        Defense           & 150.078 & $< 0.001$ & 0.158 \\
        Health            & 124.456 & $< 0.001$ & 0.144 \\
        Public Lands      & 88.759  & $< 0.001$ & 0.122 \\
        Social Welfare    & 67.716  & $< 0.001$ & 0.106 \\
        \hline
        \multicolumn{4}{l}{\textbf{Low Disagreement} ($0.05 \leq V < 0.10$)} \\
        \hline
        Environment         & 57.390  & $< 0.001$ & 0.098 \\
        Culture             & 51.718  & $< 0.001$ & 0.093 \\
        Housing             & 46.004  & $< 0.001$ & 0.088 \\
        International Affairs & 40.186 & $< 0.001$ & 0.082 \\
        Transportation      & 28.418  & $< 0.001$ & 0.069 \\
        Labor               & 24.268  & $< 0.001$ & 0.064 \\
        Civil Rights        & 22.185  & $< 0.001$ & 0.061 \\
        Technology          & 20.189  & $< 0.001$ & 0.058 \\
        Immigration         & 18.284  & $< 0.001$ & 0.055 \\
        Foreign Trade       & 16.860  & $< 0.001$ & 0.053 \\
        Energy              & 15.991  & $< 0.001$ & 0.052 \\
        \hline
        \multicolumn{4}{l}{\textbf{Very Low Disagreement} ($V < 0.05$)} \\
        \hline
        Agriculture         & 14.160  & 0.00270  & 0.049 \\
        Education           & 8.334   & 0.0396   & 0.037 \\
        Interstate Commerce & 0.600   & 0.896    & 0.010 \\
        \hline
    \end{tabular}%
    \caption{Chi-squared and Cramer's V results grouped by disagreement level across intervention methods.}
    \label{tab:chi2_classwise_grouped}
\end{table}

\paragraph{Disagreement Across Intervention Methods by Policy Class.}
We evaluated classification disagreement across intervention methods using chi-squared tests and Cramér’s V for each policy class. Results revealed \textbf{high disagreement} ($V \geq 0.20$) in classes such as \textit{Government Operations} and \textit{Law and Crime}, suggesting that methods interpret these categories differently, likely due to their conceptual breadth or overlap with other domains.

\textbf{Moderate disagreement} ($0.10 \leq V < 0.20$) was observed in economically sensitive or complex policy areas like \textit{Domestic Commerce}, \textit{Macroeconomics}, and \textit{Defense}. These may reflect nuanced interpretive boundaries between classes or varying priors introduced by prompt designs.

\textbf{Low disagreement} ($0.05 \leq V < 0.10$) was found in more narrowly defined or consistently treated classes, including \textit{Environment}, \textit{Culture}, and \textit{Transportation}. Meanwhile, \textbf{very low disagreement} ($V < 0.05$), especially in \textit{Interstate Commerce} and \textit{Education}, indicates robust cross-method consistency.

\begin{figure}[h]
    \centering
    \includegraphics[width=0.90\textwidth]{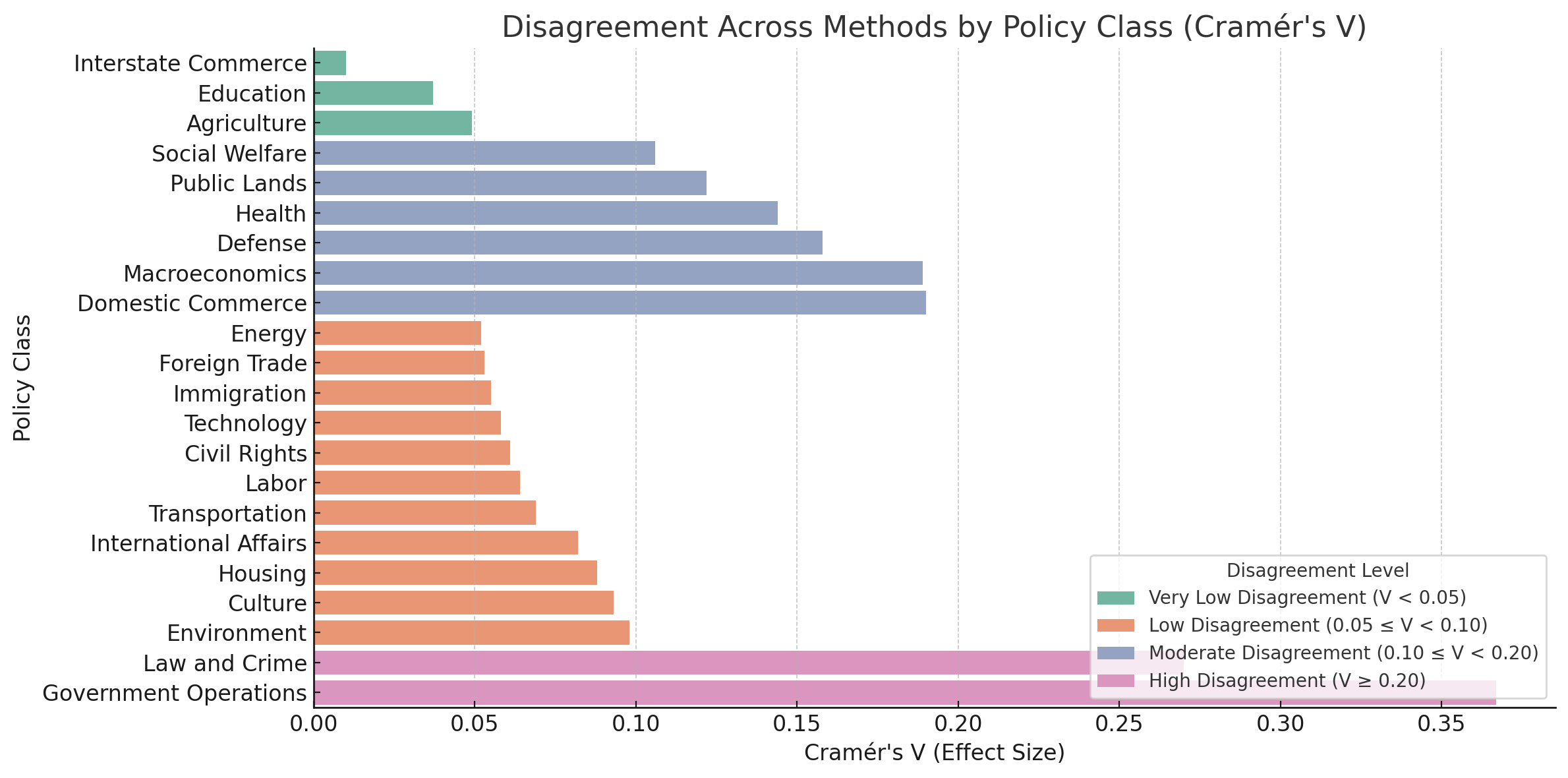}
    \caption{Cramér's V effect size per policy class, grouped by disagreement level. Higher values indicate greater divergence in classification outcomes across intervention methods.}
    \label{fig:major_class_disagreement}
\end{figure}

\subsection{Classification Performance by Method}
The interactive prompting method outperformed all other intervention strategies across all classification metrics. It achieved the highest accuracy (0.775), precision (0.685), recall (0.742), and both macro and weighted F1-scores (0.699 and 0.755, respectively). Agreement measures were similarly strong, with Cohen’s $\kappa$ = 0.744, Krippendorff’s $\alpha$ = 0.746, and Spearman’s $\rho$ = 0.732. In contrast, the zero-shot and few-shot methods performed substantially worse across the board, with F1-scores below 0.26 and interrater agreement metrics near or below 0.30. The definitions-based method showed modest improvement over few-shot prompting but did not approach the performance of interactive prompting. These results suggest that allowing models to iteratively refine their decisions via interaction can significantly enhance classification reliability and alignment with ground truth.

\begin{table}[H]
 \resizebox{\textwidth}{!}{%
\centering
\small
\begin{tabular}{lcccccccc}
\hline
\textbf{Method} & \textbf{Accuracy} & \textbf{Precision} & \textbf{Recall} & \textbf{F1 (Macro)} & \textbf{F1 (Weighted)} & \textbf{Kappa} & \textbf{Alpha} & \textbf{Spearman} \\
\hline
Zero-shot    & 0.501 & 0.353 & 0.360 & 0.333 & 0.423 & 0.353 & 0.326 & 0.278 \\
Few-shot     & 0.541 & 0.407 & 0.417 & 0.386 & 0.532 & 0.457 & 0.458 & 0.380 \\
Definitions  & 0.550 & 0.395 & 0.401 & 0.374 & 0.503 & 0.445 & 0.442 & 0.397 \\
Interactive  & 0.775 & 0.685 & 0.742 & 0.699 & 0.755 & 0.744 & 0.746 & 0.732 \\
\hline
\end{tabular}
}
\caption{Classification performance metrics across intervention methods. Interactive prompting yields the highest agreement and predictive performance across all metrics.}
\label{tab:intervention_performance}
\end{table}

\subsubsection{Classifcation Metrics by Major Class}

\begin{table}[h]
 \resizebox{0.85\textwidth}{!}{%
    \centering
    \small
    \begin{tabular}{lcccc}
        \hline
        \textbf{Class} & \textbf{Precision} & \textbf{Recall} & \textbf{F1-score} & \textbf{Support} \\
        \hline
        Energy                  & 0.967 & 1.000 & 0.983 & 29 \\
        Immigration             & 0.938 & 1.000 & 0.968 & 30 \\
        Law and Crime           & 0.908 & 0.834 & 0.869 & 415 \\
        Agriculture             & 0.913 & 0.724 & 0.808 & 29 \\
        Civil Rights            & 0.780 & 0.784 & 0.782 & 204 \\
        Housing                 & 0.793 & 0.767 & 0.780 & 30 \\
        Education               & 0.875 & 0.700 & 0.778 & 30 \\
        Environment             & 0.703 & 0.867 & 0.776 & 30 \\
        Domestic Commerce       & 0.775 & 0.757 & 0.766 & 177 \\
        Labor                   & 0.689 & 0.857 & 0.764 & 119 \\
        Social Welfare          & 0.667 & 0.889 & 0.762 & 27 \\
        Technology              & 0.870 & 0.667 & 0.755 & 30 \\
        Transportation          & 0.729 & 0.717 & 0.723 & 60 \\
        International Affairs   & 0.731 & 0.679 & 0.704 & 28 \\
        Public Lands            & 0.645 & 0.769 & 0.702 & 26 \\
        Culture                 & 0.750 & 0.621 & 0.679 & 29 \\
        Defense                 & 0.704 & 0.655 & 0.679 & 29 \\
        Health                  & 0.656 & 0.700 & 0.677 & 30 \\
        Macroeconomics          & 0.629 & 0.733 & 0.677 & 30 \\
        Government Operations   & 0.634 & 0.670 & 0.652 & 88 \\
        Foreign Trade           & 0.789 & 0.500 & 0.612 & 30 \\
        \hline
    \end{tabular}
    }
    \caption{Classification metrics for all major policy classes under the interactive intervention, sorted by F1-score.}
    \label{tab:all_class_metrics}
\end{table}

\paragraph{Per-Class Performance under the Interactive Intervention.}
Table~\ref{tab:all_class_metrics} shows the top 15 policy classes by F1-score when using interactive prompting. Classes such as \textit{Energy}, \textit{Immigration}, and \textit{Law and Crime} achieve the highest F1 values (0.869--0.983), reflecting both high precision and recall. In contrast, domains like \textit{Public Lands} and \textit{International Affairs} see more moderate F1-scores (0.702--0.704). This performance pattern suggests that interactive prompting excels in classes with clearer decision boundaries or well-defined keyword sets, while more context-dependent areas retain moderate misclassification. Overall, the substantial F1 gains for these top classes underscore the efficacy of iterative refinement in accurately capturing complex policy topics.

\begin{figure}[h]
    \centering
    \includegraphics[width=0.95\textwidth]{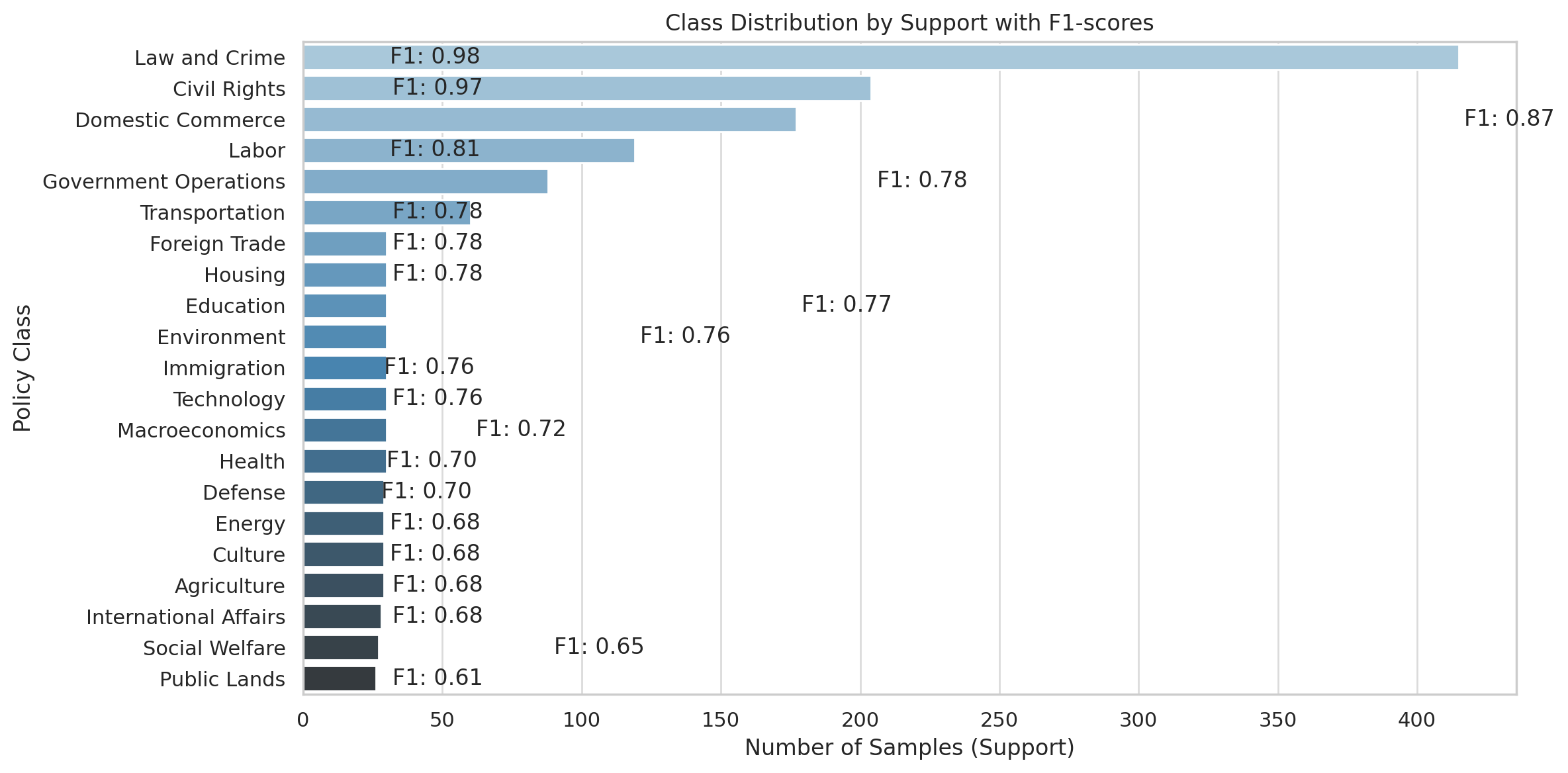}
    \caption{Distribution of policy classes by sample count (support), annotated with F1-scores under the interactive intervention. While some classes with low support (e.g., \textit{Energy}, \textit{Immigration}) achieve near-perfect F1-scores, others with higher support (e.g., \textit{Government Operations}, \textit{Foreign Trade}) show comparatively weaker performance. This suggests that F1-score is not solely driven by frequency and reflects the model’s sensitivity to class semantics.}
    \label{fig:class_f1_distribution}
\end{figure}

\subsubsection{Classification Metrics by Subclass}

\begin{table}[H]
\centering
\footnotesize
\begin{tabular}{>{\raggedright\arraybackslash}p{4.5cm}rrrr}
\hline
\textbf{Subclass} & \textbf{Precision} & \textbf{Recall} & \textbf{F1-score} & \textbf{Support} \\
\hline
General Technology        & 1.00 & 1.00 & 1.00 & 5 \\
Natural Gas \& Oil        & 0.85 & 0.78 & 0.81 & 134 \\
Immigration               & 0.73 & 0.77 & 0.75 & 195 \\
Copyrights and Patents    & 0.87 & 0.62 & 0.72 & 170 \\
Bankruptcy                & 0.57 & 0.91 & 0.70 & 120 \\
Indigenous Affairs        & 0.57 & 0.89 & 0.70 & 169 \\
Securities \& Commodities & 0.70 & 0.71 & 0.70 & 160 \\
Drug Coverage and Cost    & 1.00 & 0.50 & 0.67 & 8 \\
Military Procurement      & 1.00 & 0.50 & 0.67 & 2 \\
Minority Discrimination   & 0.65 & 0.65 & 0.65 & 329 \\
Political Campaigns       & 0.61 & 0.63 & 0.62 & 153 \\
Banking                   & 0.66 & 0.51 & 0.57 & 81 \\
Corporate Management      & 0.50 & 0.65 & 0.57 & 395 \\
General Public Lands      & 1.00 & 0.40 & 0.57 & 10 \\
Labor Unions              & 0.42 & 0.87 & 0.57 & 294 \\
Subsidies to Farmers      & 1.00 & 0.40 & 0.57 & 5 \\
Tax Administration        & 0.48 & 0.67 & 0.56 & 190 \\
Criminal \& Civil Code    & 0.37 & 0.93 & 0.53 & 1557 \\
Prisons                   & 0.77 & 0.41 & 0.53 & 123 \\
Telecommunications        & 1.00 & 0.36 & 0.53 & 50 \\
Appointments              & 1.00 & 0.33 & 0.50 & 15 \\
Interest Rates            & 1.00 & 0.33 & 0.50 & 3 \\
General Energy            & 0.63 & 0.38 & 0.48 & 13 \\
Drinking Water            & 0.44 & 0.47 & 0.46 & 17 \\
Low-Income Assistance     & 0.88 & 0.31 & 0.46 & 45 \\
Family Issues             & 0.45 & 0.45 & 0.45 & 121 \\
Claims against Military   & 0.57 & 0.36 & 0.44 & 11 \\
Other Commerce            & 1.00 & 0.29 & 0.44 & 7 \\
Right to Privacy          & 0.63 & 0.32 & 0.43 & 143 \\
Branch Relations          & 0.58 & 0.33 & 0.42 & 57 \\
Freedom of Speech         & 0.56 & 0.32 & 0.41 & 400 \\
\hline
\end{tabular}
\caption{Precision, recall, F1-score, and support for each subclass, sorted by F1-score.}
\label{tab:subclass_metrics_sorted}
\end{table}

To assess baseline performance prior to introducing intervention methods, we computed classification metrics for each of the 220 subclasses. Table~\ref{tab:subclass_metrics_sorted} presents the top-performing subclasses ranked by F1-score. While several high-performing categories exhibited strong F1 scores and substantial support, such as \textit{Natural Gas \& Oil} ($F_1 = 0.81$, $n = 134$) and \textit{Immigration} ($F_1 = 0.75$, $n = 195$), others achieved similarly high F1 scores despite limited sample sizes. For instance, \textit{General Technology} achieved perfect precision, recall, and F1-score ($F_1 = 1.00$), but had only 5 labeled instances. Other subclasses like \textit{Military Procurement}, \textit{Subsidies to Farmers}, and \textit{Interest Rates} also reported high F1 scores ($F_1 = 0.50$–$0.67$) with support below $n = 10$, limiting the generalizability of these metrics. These results represent ChatGPT's unassisted ability to assign subclass labels based solely on case summaries, offering a performance baseline against which the effectiveness of intervention strategies can be evaluated.

\section{Discussion}

Our results demonstrate that structured prompting can significantly enhance ChatGPT’s ability to perform deductive qualitative coding. However, this performance remains sensitive to input complexity, code ambiguity, and instruction decay, underscoring the importance of continued refinement in intervention design and task structuring. 

Throughout the coding tasks we found that, without sufficient steering, ChatGPT tended to characterize the legal summaries in terms of the most general or broadest category. Because CAP major labels exhibit semantic overlap, the models struggled to determine which conceptual element should guide label assignment. We found these difficulties to be attributable either to the complexity of the summary or the counterintuitive label assigned by human coders. This was evidenced with major label disagreement between interventions being highest in semantically expansive categories like Law and Crime and Government Operations, and lowest in narrower-scoped categories like Agriculture, Education, and Interstate Commerce, suggesting that ChatGPT may scale more effectively when applying specific or low-level codes. Further, longer summaries with several thematic loci tended to confuse the model by presenting it with competing themes without a criterion for thematic precedence. In contrast, shorter summaries of several sentences in length facilitated quicker identification of the correct class. This suggests that effective integration of LLMs in deductive coding may require chunking qualitative data into discrete semantic units to support accurate classification. 

While differences in performance between the zero-shot, few-shot, and definition-based interventions were relatively modest, the step-by-step decomposition intervention demonstrated a substantial performance improvement, suggesting that level-of-detail feedback can steer the model toward task-specific application of the coding scheme. Since each intervention was conducted in a separate chat, the consistency of results across chats suggests that the step-by-step intervention achieves both validity and high inter-rater reliability, approximating the performance of a trained human coder. At the same time, we observed signs of performance degradation over repeated classification instances typically between the 30th and 40th summary, consistent with instruction decay in generative LLMs. These findings suggest that when incorporating generative LLMs for scalable deductive coding tasks, input batches should be limited to the performance degradation threshold. Robustly quantifying this threshold represents a promising direction for future research. 

Finally, benchmarking against an opaque, purely discriminative baseline such as RoBERTa yielded the highest quantitative scores (accuracy=0.79, weighted F1=0.79, =0.75) relative to the step-by-step ChatGPT intervention (accuracy=0.775, weighted F1=0.755, =0.744), a comparison intended to establish an upper bound on metric-focused performance. However, ChatGPT’s generative reasoning chain offered transparent, auditable justifications that streamline member checking, collaborative re-coding, and consensus-building, thereby offsetting its marginal deficit in performance. We now proceed to answer the research questions that we posted at the outset of our study.

\paragraph{Baseline Model Comparison.} In the baseline prediction task, ChatGPT performed comparably to a Bidirectional Long Short-Term Memory (BiLSTM) model. For the major class, ChatGPT achieved an accuracy of 0.57 and a Cohen’s $\kappa$ of 0.46, whereas the BiLSTM reached an accuracy of 0.59 and a $\kappa$ of 0.55. A custom-trained RoBERTa model outperformed both, attaining an accuracy of 0.79 and a Cohen’s $\kappa$ of 0.75. For the subclass task, ChatGPT again outperformed the BiLSTM, with an accuracy of 0.46 and a $\kappa$ of 0.41, compared to the BiLSTM’s accuracy of 0.21 and $\kappa$ of 0.244. The RoBERTa model achieved the highest performance for subclass prediction as well, with an accuracy of 0.65 and a $\kappa$ of 0.63. However, despite their performance advantages, both the BiLSTM and RoBERTa models operate as black boxes and do not provide interpretable outputs. This limits their utility for social scientists who require transparent, human-interpretable and human-interoperable justifications for classification decisions, particularly in the context of applying structured coding schemes.

\paragraph{Convergent and discriminant validity across interventions.} As a result we constructed four distinct intervention methods with the aim of improving ChatGPT classification performance metrics while also maintaining the interpretability of the classification decisions. Our aim was twofold: to determine whether intervention methods displayed statistically significant differences, and whether they internally cohered. We found that zero-shot, few-shot, definition-based, and step-by-step reasoning constitute distinct intervention methods. We found that zero-shot and few-shot interventions exhibited greater within-method coherence, as evidenced by fewer statistically significant pairwise differences, particularly after Bonferroni and FDR corrections. In contrast, definition-based and step-by-step reasoning (interactive) methods showed more internal variation, with a higher number of significant comparisons across samples, suggesting greater sensitivity to sampling differences or procedural variation. These results suggest that apprising ChatGPT with more contextual information produces more volatile results. 

Conversely, the strongest divergence was observed between the \textit{interactive} and \textit{zero-shot} methods, which produced the highest mean chi-squared value ($\bar{\chi}^2 = 314.27$) and the greatest proportion of statistically significant tests (30/30 after Bonferroni correction). In contrast, the \textit{definitions} and \textit{few-shot} methods exhibited the weakest divergence, with the lowest mean chi-squared value ($\bar{\chi}^2 = 148.65$) and the fewest significant results (18/30 after Bonferroni correction). These results suggest that the degree of disagreement between methods varies substantially, even though overall discriminant validity remains strong across all comparisons. Broadly, the intervention methods show greater discriminant than convergent validity, though they all meet the sufficiency threshold of statistical significance. 

\paragraph{Intervention Performance Comparison.} The differences in performance between the zero-shot, few-shot, and definition-based interventions were relatively modest. However, accuracy increased incrementally across these methods: from zero-shot ($0.50$), to few-shot ($0.54$), to definition-based prompting ($0.55$). In contrast, the step-by-step reasoning intervention demonstrated a substantial performance improvement, achieving an accuracy of $0.77$, a weighted F1-score of $0.75$, Cohen’s $\kappa$ of $0.744$, and Krippendorff’s $\alpha$ of $0.75$. These reliability coefficients approach the threshold for substantial agreement, thereby lending support to the validity of the coding scheme. Notably, since each classification task was conducted in a separate chat instance, each session can be conceptually treated as a distinct rater. The consistency of results across chats suggests that the step-by-step reasoning intervention achieves both validity and high inter-rater reliability, approximating the performance of a trained human coder. 

\paragraph{Interpretation of Interactive Reasoning.} ChatGPT performance showed variation across models. Throughout the interactive intervention we confined our coding tasks to ChatGPT 4o, 4.5, o1, and o1 mini. These state-of-the-art (SOTA) models excel at step-by-step reasoning and complex analytic tasks. During the deductive coding task, we found that without sufficient steering, these models tend to characterize the legal summaries in terms of the most general or broadest category. Because CAP major classes exhibit semantic overlap, the models showed considerable difficulty deciding which conceptual element should determine the label assignment. However, these difficulties were sometimes attributable to the complexity of the summary and other times the counterintuitive ground-truth label assigned by human reasoners. 

The models tended to assign the \textit{Law and Crime} category more frequently than human coders. For this reason, the models were explicitly instructed to only apply \textit{Law and Crime} if the other categories were not explicitly applicable. Common mislabelings occurred between \textit{Labor} and \textit{Civil Rights}, \textit{Labor} and \textit{Culture}, \textit{Agriculture} and \textit{Public Lands}, \textit{Agriculture} and \textit{Environment}, \textit{Domestic Commerce} and \textit{Transportation}, \textit{International Affairs} and \textit{Foreign Trade}, \textit{Housing} and \textit{Environment}, \textit{Social Welfare }and \textit{Government Operations}, \textit{Health} and \textit{Labor}, \textit{Health} and \textit{Environment}, \textit{Domestic Commerce} and \textit{Macroeconomics}. 

To illustrate, the summaries indicate that the human coders used certain keywords or conceptual categories as indicators of a particular label. For example, all tax related matters were assigned the \textit{Macroeconomics} label. However, since the model is not privy to the weight of the term tax, it often labeled the summary with the \textit{Government Operations }or \textit{Domestic Commerce} label instead. Another salient example pertains to how the human coders applied the \textit{Culture} label. In most cases, the \textit{Culture} label was assigned to labor legal cases where the defendants or plaintiffs operated in the entertainment industry. Because the model was not privy to the weight the human coders placed on the entertainment industry keywords, the model was more likely to ascribe the label \textit{Labor}, since the legal case concerned general labor matters. Even when corrected, the model sometimes would display recalcitrance and resist revising its classification unless provided with explicit evidence from the summary that a competing label was applicable. The positive aspect of these mislabeling cases was that the model in the vast majority of cases chose a label that was semantically supported by the legal summary. Below, we provide five examples of mislabelings along with the model rationale for choosing the label. 

\paragraph{ChatGPT Mislabeling Example:1.}
\vspace{1em} 
In the example below, the human coded label is \textit{Foreign Trade}, but ChatGPT has reasonably chosen \textit{Transportation}:

\begin{figure}[htbp]
    \centering
    \includegraphics[width=\textwidth]{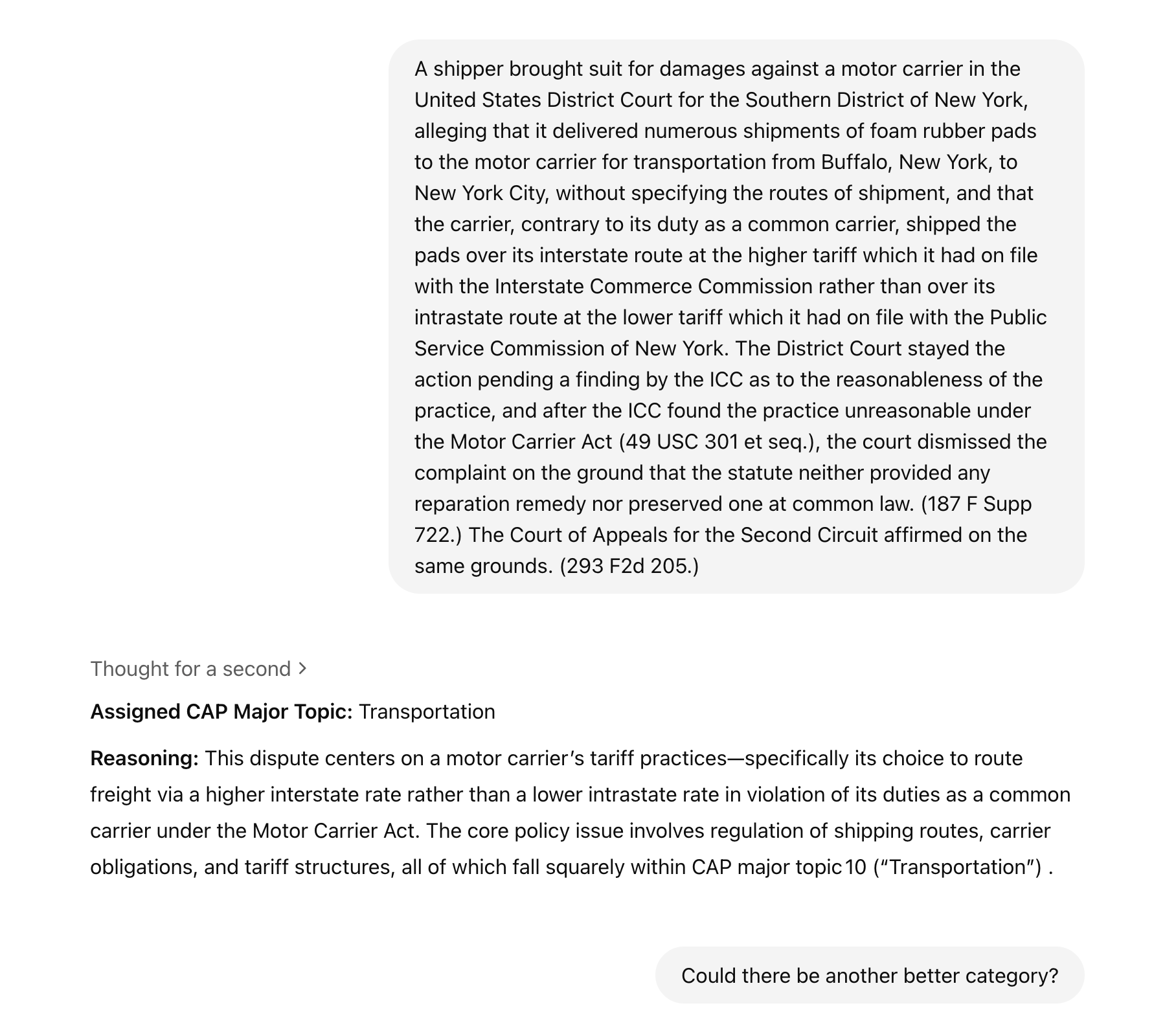}
    \caption{ChatGPT mislabels summary for Foreign Trade.}
    \label{fig:ChatGPT_classified1}
\end{figure}

\begin{figure}[H]
    \centering
    \includegraphics[width=\textwidth]{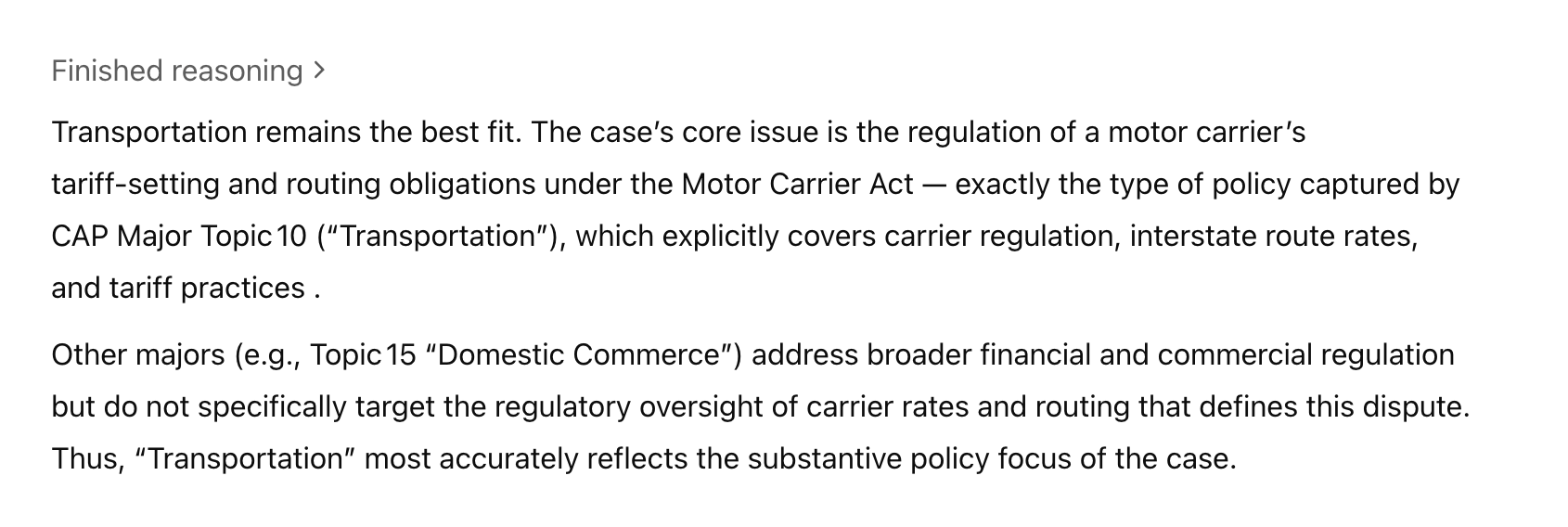}
    \caption{ChatGPT sticks to its reasoning on Transportation label.}
    \label{fig:ChatGPT_classified2}
\end{figure}

\paragraph{ChatGPT Mislabeling Example:2.}
\vspace{1em} 
In the example below, the human coder label is \textit{Energy}, but ChatGPT has reasonably chosen \textit{Labor}:

\begin{figure}[H]
    \centering
    \includegraphics[width=\textwidth]{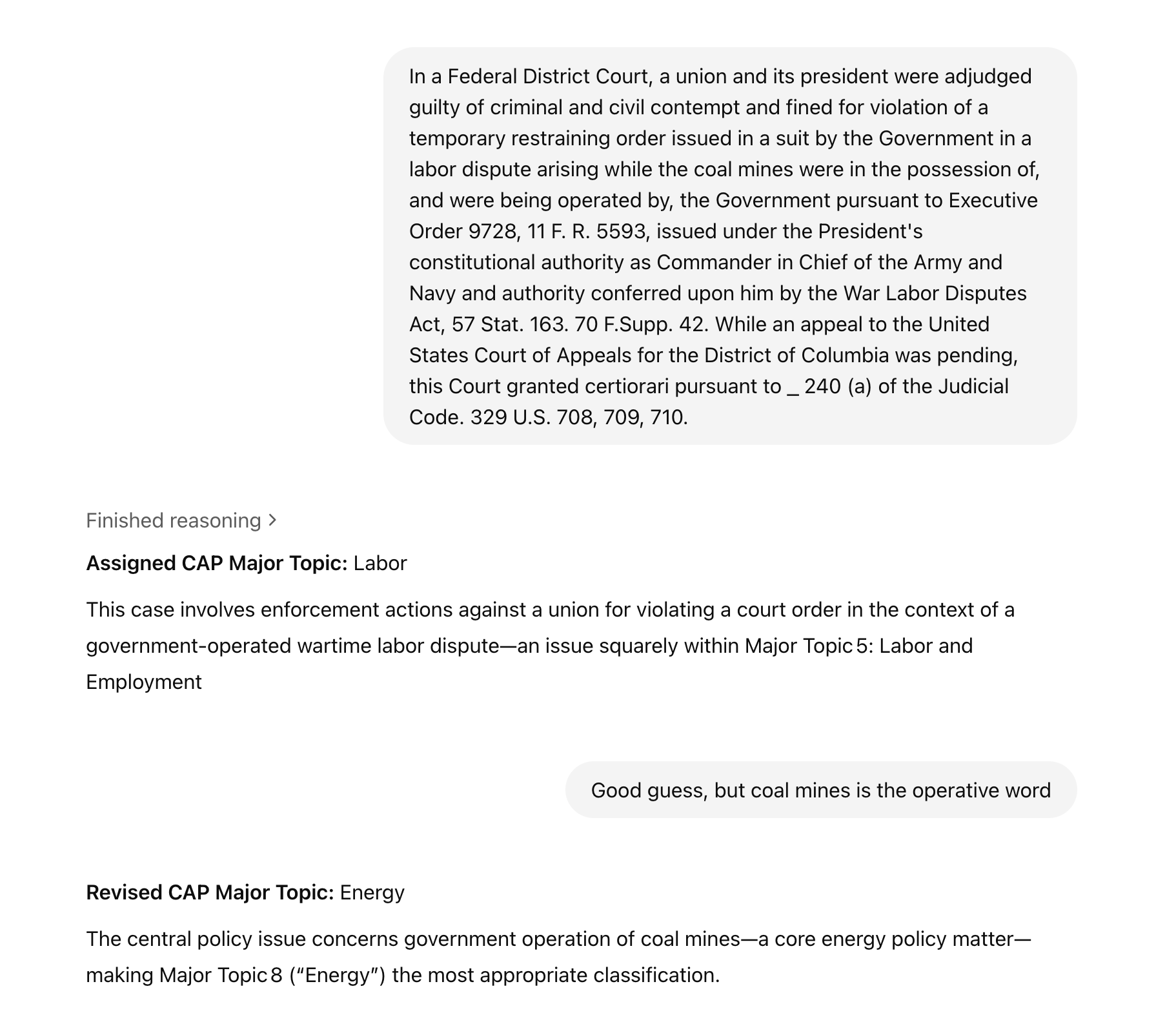}
    \caption{ChatGPT revises original classification upon prompt from Label to Energy.}
    \label{fig:ChatGPT_classified3}
\end{figure}

\paragraph{ChatGPT Mislabeling Example:3.}
\vspace{1em} 
In the example below, the human coder has appropriately labeled the legal summary \textit{Government Operations} since the case is against the Federal Election Commission. However, ChatGPT labeled the summary \textit{Civil Rights}, since the case concerned electioneering and the constitutionality of free speech. Human coders tend to assign \textit{Government Operations} any election-related legal cases, whereas ChatGPT identified the more general issue of free speech as the overriding theme: 

\begin{figure}[H]
    \centering
    \includegraphics[width=\textwidth]{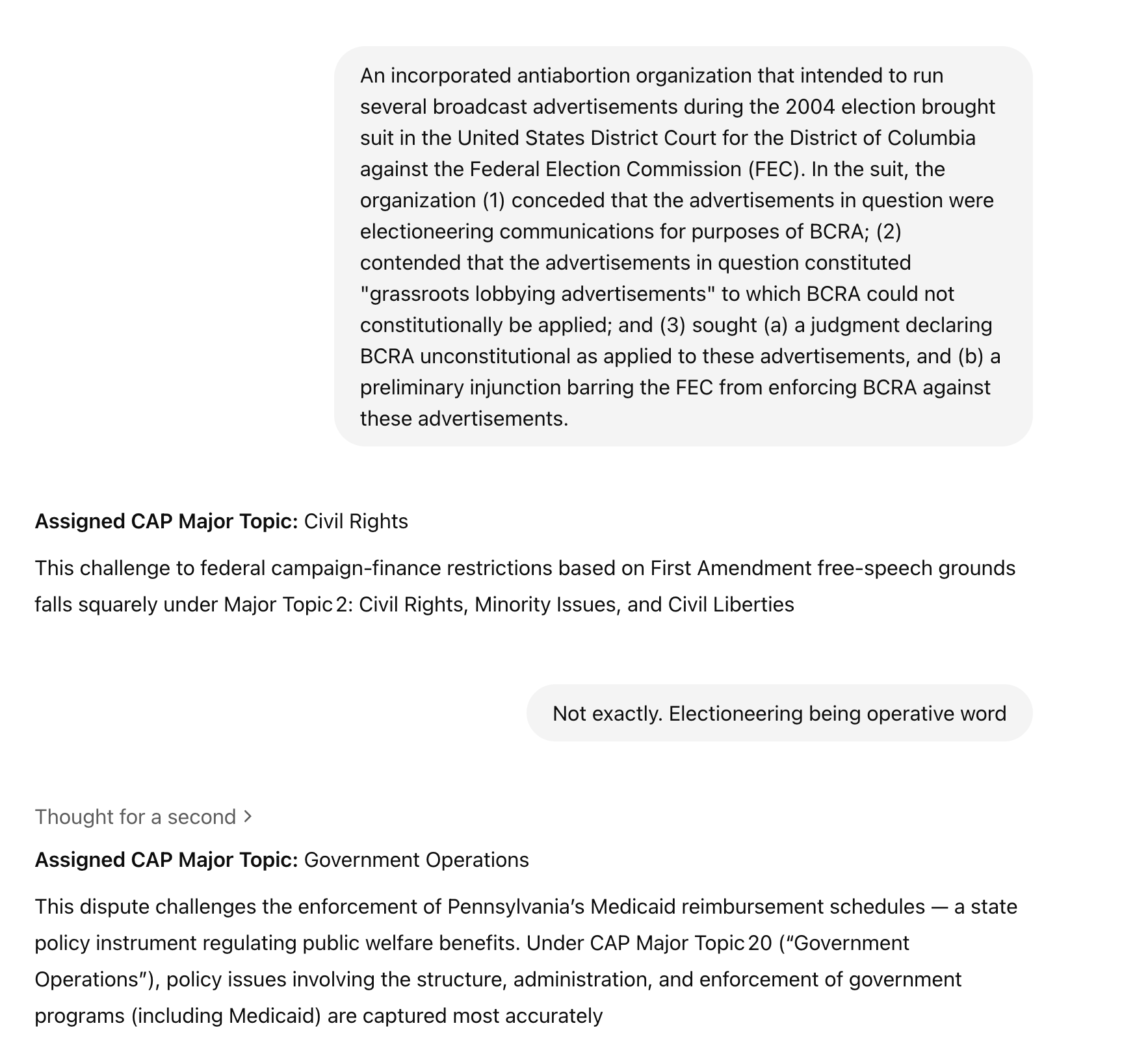}
    \caption{ChatGPT mislabels Government Operations as Civil Rights.}
    \label{fig:ChatGPT_classified4}
\end{figure}

\paragraph{ChatGPT Mislabeling Example:4.}
\vspace{1em} 
In the example below, the human coder has assigned the label \textit{Agriculture} but ChatGPT identified \textit{Transportation}: 

\begin{figure}[H]
    \centering
    \includegraphics[width=\textwidth]{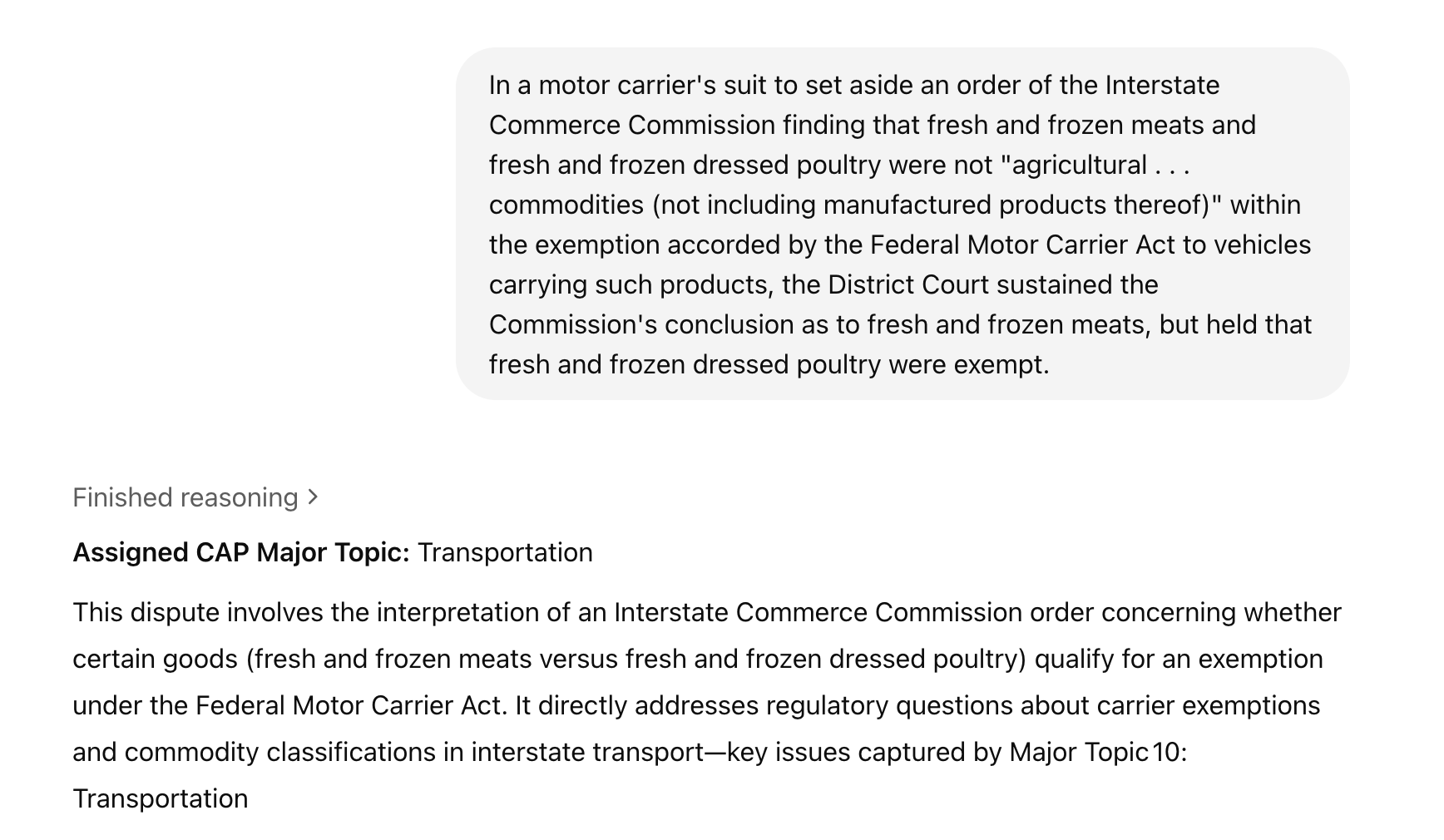}
    \caption{ChatGPT mislabels Agriculture as Transportation.}
    \label{fig:ChatGPT_classified5}
\end{figure}

\paragraph{ChatGPT Mislabeling Example:5.}
\vspace{1em} 
In the example below, the human coder has assigned the label \textit{Civil Rights} but ChatGPT identified \textit{Public Lands}: 

\begin{figure}[H]
    \centering
    \includegraphics[width=\textwidth]{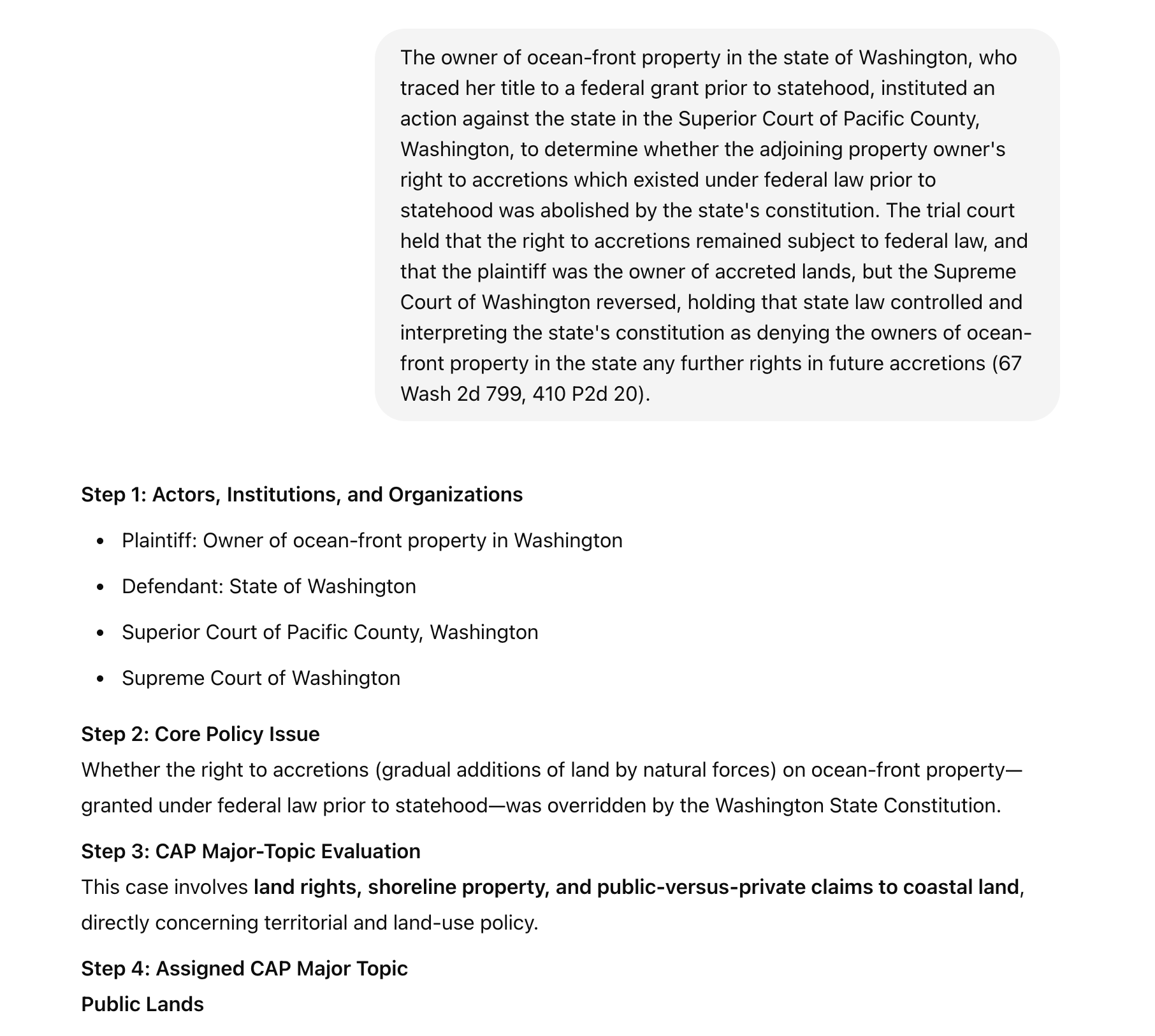}
    \caption{ChatGPT mislabels Civil Rights as Public Lands.}
    \label{fig:ChatGPT_classified5}
\end{figure}

Broadly, ChatGPT displayed sensitivity to the complexity of the summary in accurately assinging the label. Longer summaries that contained several thematic loci tended to confuse the model by presenting it with several competing themes without explicit indication as to which theme should take conceptual precedence. Shorter summaries of several sentences in length, by contrast, resulted in quicker identification of the correct class. Over the course of a classification sample of 50 summaries total, ChatGPT also displayed a tendency to become less sharp and default to more generic labels. This decline in quality typically occurred approximately between the 30th and 40th summary. We now proceed to answer the research questions that we posed at the outset of our study. 

\vspace{1ex}

\textbf{Q1:} \textbf{What is the baseline reliability of ChatGPT in deductive coding?}
\vspace{1ex}
Our findings reveal that with the right prompting strategy and in-depth knowledge of the coding task ChatGPT can be steered toward reaching a range of accuracy between 0.7-.08 in deductive coding tasks. We found that our \textit{Step-by-Step Task Decomposition} prompting strategy can reliably attain high performance metrics across a large number of samples. These findings can be extrapolated with reasonable confidence to comparable chatbots with similar benchmarks. 

\vspace{1ex}
\textbf{Q2:} \textbf{Are there distinct prompt engineering interventions and coding pipelines that improve benchmarks?}
We found that our fourth intervention strategy, \textit{Step-by-Step Task Decomposition}, produced the most accurate and reliable results. This approach involved explicitly instructing ChatGPT to articulate its reasoning process, while also providing a concise description of the classification task and relevant contextual information. The intervention achieved an accuracy of $0.76$, with inter-coder agreement metrics of Cohen's $\kappa = 0.74$, Krippendorff's $\alpha = 0.76$, and Spearman's rank correlation $\rho = 0.73$. These values fall within the threshold of \textit{substantial agreement}, demonstrating that the intervention effectively enhances ChatGPT's ability to perform deductive qualitative coding with reliability approaching that of trained human coders.

\vspace{1ex}
\textbf{Q3: Can a generative chatbot serve as a qualitative coding assistant?} Based on the results of our study, we provisionally answer this question in the affirmative, while acknowledging that further studies are needed to generalize these findings beyond our specific task. Our classification task, labeling U.S. Supreme Court case summaries, is relatively constrained and domain-specific. However, the reasoning capabilities exhibited by ChatGPT in this context suggest that its potential applicability extends to a broader range of deductive and inductive qualitative coding tasks.

We conclude that ChatGPT and comparable autoregressive models can significantly streamline deductive qualitative coding. Although we did not fully automate the reasoning process due to computational and budgetary constraints, our intervention strategy, \textit{Step-by-Step Task Decomposition}, is inherently automatable. This opens the door for scaling qualitative coding across large corpora. Whether researchers choose to treat ChatGPT as a full-fledged coder or as an assistive tool to enhance consistency and validity will depend on the specific research design and coding requirements. Nonetheless, our findings indicate that either role is justifiable given the current capabilities of state-of-the-art language models.

\section{Limitations}
Our findings should be put in context of experimental limitations and potential confounds that limit the generalizability of our results. 

A potential limitation of our study stems from intrinsic limitations within the dataset. We are unable to independently validate the validity and reliability of the CAP Codebook or the accuracy of the labels. Consequently, we assumed accuracy and treated the classified datasets provided by the CAP Project as ground truth and the CAP Codebook as a valid and reliable instrument. In light of this assumption, it is important to consider the potential for human error and the rationale for the application of the scheme. While we treated major classes as a mutually exclusive and exhaustive set, the assumption of mutual exclusivity and exhaustion is highly interpretative in the context in legal areas. This poses a major semantic challenge in establishing the validity and reliability of the classification scheme. These semantic challenges were evident throughout the classification tasks we conducted with ChatGPT where a given case summary was often compatible with a different label than the assigned ground truth label. In some cases, the rationale that ChatGPT provided for a given label was more strongly supported by the textual evidence than the ground truth label. The incidence of these occurrences was relatively frequent, indicating a wide window of interpretation in the application of the codebook. The issue of interpretability should be taken into consideration in the context of ChatGPT's classification performance with respect to accuracy, precision, and recall. 

\subsection{Confounds}
Some potential confounds include inherent ambiguity in the case summaries, the inability to independently assess the accuracy of the ground truth labels, sensitivity to prompt variations (though these were controlled across methods), and inherent propensity for variance within the model. Broadly speaking, the input independent variable, namely the Supreme Court Case summaries, exhibited variation with respect to length, clarity, amount of legal jargon and legalese, and semantic clarity. These variables likely had an effect on the ability of the model to infer the correct label. Cognate to summary ambiguity, the CAP coding scheme itself contains ambiguity. As already noted, the set of major classes do not strictly constitute a mutually exclusive and exhaustive set likely because the policy topic is not amenable to this kind of logically strict partition. As a result, even with explicit definitions, the model was likely to confuse some major classes for each other. A potential method to avoid these mislabelings is to supply the model with explicit rules that deterministically produce label outputs from the right textual cues. This suggests a path for a future study that requires liaising with CAP researchers. 

\section{Conclusion}
In light of recent research exploring the integration of large language models (LLMs) into qualitative coding workflows, our study addresses persistent gaps in applying AI chatbots to automated and semi-automated deductive coding tasks. While the preponderance of recent work demonstrates encouraging results with ChatGPT and similar models in inductive coding, where task structures are more flexible and interpretability more permissible, there remains a scarcity of studies focused on structured, deductive coding. Our aim was to evaluate whether an intervention strategy could guide ChatGPT to achieve acceptable classification accuracy and levels of intercoder agreement aligned with standards in social science research. To this end, we selected a complex human-coded classification scheme with established validity and tested a \textit{Step-by-Step Decomposition} prompt strategy in a case-by-case coding setup. This intervention produced consistently high agreement scores with human-coded data. While we did not perform a regression analysis to isolate sources of variance, we hypothesize that remaining discrepancies stem from the inherent ambiguity of some case summaries and the broad latitude for semantic interpretation in legal language. Nevertheless, our intervention achieved Cohen’s $\kappa$ and Krippendorff’s $\alpha$ values that fall within the widely accepted \textit{substantial agreement} range in both industry and academic research as benchmarks for reliable classification. These findings suggest that, with minimal scaffolding, LLMs can meet reliability thresholds sufficient for integration into rigorous qualitative workflows.


\begin{thebibliography}{10}

\bibitem{baumgartner2019}
Frank~R. Baumgartner, Christian Breunig, and Emiliano Grossman.
\newblock The comparative agendas project: Intellectual roots and current developments.
\newblock \url{https://kops.uni-konstanz.de/entities/publication/ef02baf4-c9e9-4dd4-a166-0b04960b9c7d}, 2019.
\newblock Working paper.

\bibitem{bevan2020}
Shaun Bevan and Anna~M. Palau.
\newblock The comparative agendas project in latin america: Data and coding.
\newblock {\em Revista de Administración Pública}, 2020.
\newblock Published August.

\bibitem{bijker2024}
R.~Bijker, S.~S. Merkouris, N.~A. Dowling, and S.~N. Rodda.
\newblock Chatgpt for automated qualitative research: Content analysis.
\newblock {\em Journal of Medical Internet Research}, 26:e59050, 2024.
\newblock Published July 25.

\bibitem{Bringer2006}
J.~D. Bringer, L.~H. Johnston, and C.~H. Brackenridge.
\newblock Using computer-assisted qualitative data analysis software to develop a grounded theory project.
\newblock {\em Field Methods}, 18(3):245--266, 2006.
\newblock Original work published 2006.

\bibitem{chaturvedi2015}
Santosh R. B.~H. Chaturvedi and R.~C. Shweta.
\newblock Evaluation of inter-rater agreement and inter-rater reliability for observational data: An overview of concepts and methods.
\newblock {\em Journal of the Indian Academy of Applied Psychology}, 41(3):20--27, 2015.

\bibitem{chung2020}
Chia-Jung Chung, J.~Patrick Biddix, and Han~Woo Park.
\newblock Using digital technology to address confirmability and scalability in thematic analysis of participant-provided data.
\newblock {\em The Qualitative Report}, 25(9):3298--3311, 2020.

\bibitem{cohen1960}
Jacob Cohen.
\newblock A coefficient of agreement for nominal scales.
\newblock {\em Educational and Psychological Measurement}, 20(1):37--46, 1960.

\bibitem{cypress2019}
Brigitte~S. Cypress.
\newblock Data analysis software in qualitative research: Preconceptions, expectations, and adoption.
\newblock {\em Dimensions of Critical Care Nursing}, 38(4):213--220, 2019.
\newblock Published July/August.

\bibitem{dowding2016}
Keith Dowding, Andrew Hindmoor, and Aaron Martin.
\newblock The comparative policy agendas project: Theory, measurement and findings.
\newblock {\em Journal of Public Policy}, 36(1):3--25, 2016.

\bibitem{gao2024}
Jie Gao et~al.
\newblock Collabcoder: A lower-barrier, rigorous workflow for inductive collaborative qualitative analysis with large language models.
\newblock In {\em Proceedings of the CHI Conference on Human Factors in Computing Systems}, pages 1--29. ACM, 2024.

\bibitem{gebreegziabher2023}
Simret~Araya Gebreegziabher, Zheng Zhang, Xiaohang Tang, Yihao Meng, Elena~L. Glassman, and Toby Jia-Jun Li.
\newblock Patat: Human-ai collaborative qualitative coding with explainable interactive rule synthesis.
\newblock In {\em Proceedings of the 2023 CHI Conference on Human Factors in Computing Systems}, volume~2, pages 1--19, New York, NY, USA, 2023. ACM.

\bibitem{gwet2021}
Kilem~L. Gwet.
\newblock {\em Handbook of Inter-Rater Reliability}.
\newblock STATAXIS Publishing Company, Gaithersburg, MD, 2001.

\bibitem{hallgren2012}
Kevin~A. Hallgren.
\newblock Computing inter-rater reliability for observational data: An overview and tutorial.
\newblock {\em Tutorials in Quantitative Methods for Psychology}, 8(1):23--34, 2012.

\bibitem{krippendorff2011}
Klaus Krippendorff.
\newblock Computing krippendorff’s alpha-reliability.
\newblock \url{https://citeseerx.ist.psu.edu/document?repid=rep1&type=pdf&doi=de8e2c7b7992028cf035f8d907635de871ed627d}, 2011.
\newblock Technical report.

\bibitem{landis1977}
J.~R. Landis and G.~G. Koch.
\newblock The measurement of observer agreement for categorical data.
\newblock {\em Biometrics}, 33(1):159--174, 1977.

\bibitem{marathe2018}
Megh Marathe and Kentaro Toyama.
\newblock Semi-automated coding for qualitative research: A user-centered inquiry and initial prototypes.
\newblock In {\em Proceedings of the 2018 CHI Conference on Human Factors in Computing Systems (CHI '18)}, pages Paper 348, 1--12, New York, NY, USA, 2018. Association for Computing Machinery.

\bibitem{mcdonald2019}
Nora McDonald et~al.
\newblock Reliability and inter-rater reliability in qualitative research: Norms and guidelines for cscw and hci practice.
\newblock {\em Proceedings of the ACM on Human-Computer Interaction}, 3(CSCW):1--23, 2019.

\bibitem{mchugh2012}
Mary~L. McHugh.
\newblock Interrater reliability: The kappa statistic.
\newblock {\em Biochemia Medica (Zagreb)}, 22(3):276--282, 2012.

\bibitem{morgan2023}
David~L. Morgan.
\newblock Exploring the use of artificial intelligence for qualitative data analysis: The case of chatgpt.
\newblock {\em International Journal of Qualitative Methods}, 22, 2023.

\bibitem{nelson2018}
L.~K. Nelson, D.~Burk, M.~Knudsen, and L.~McCall.
\newblock The future of coding: A comparison of hand-coding and three types of computer-assisted text analysis methods.
\newblock {\em Sociological Methods \& Research}, 50(1):202--237, 2018.
\newblock Original work published 2021.

\bibitem{olapane2021}
Elias~C. Olapane.
\newblock An in-depth exploration on the praxis of computer-assisted qualitative data analysis software (caqdas).
\newblock {\em Journal of Humanities and Social Sciences Studies}, 2021.
\newblock Published November.

\bibitem{orr2024}
Will Orr and Kate Crawford.
\newblock The social construction of datasets: On the practices, processes, and challenges of dataset creation for machine learning.
\newblock {\em New Media \& Society}, 26(9):4955--4972, 2024.

\bibitem{cap_supreme_court}
Comparative~Agendas Project.
\newblock Supreme court cases dataset.
\newblock \url{https://www.comparativeagendas.net/datasets}, 2021.
\newblock Accessed April 2025.

\bibitem{rietz2021}
Tim Rietz and Alexander Maedche.
\newblock Cody: An ai-based system to semi-automate coding for qualitative research.
\newblock In {\em Proceedings of the 2021 CHI Conference on Human Factors in Computing Systems}, pages 1--14. ACM, 2021.

\bibitem{rodriguez2022}
Patricia Rodriguez~Espinosa et~al.
\newblock Found in translation: Reflections and lessons for qualitative research collaborations across language and culture.
\newblock {\em International Journal of Qualitative Methods}, 21, 2022.

\bibitem{saldana2015}
Johnny~M. Saldaña.
\newblock {\em The Coding Manual for Qualitative Researchers}.
\newblock SAGE Publications, London, England, 3 edition, 2015.

\bibitem{schmidt2021}
Lena Schmidt, Ayse~Nazli Finnerty~Mutlu, Rebecca Elmore, Babatunde~K. Olorisade, James Thomas, and Julian P.~T. Higgins.
\newblock Data extraction methods for systematic review (semi)automation: Update of a living systematic review.
\newblock {\em F1000Research}, 10:401, 2021.

\bibitem{sinkovics2012}
Rudolf~R. Sinkovics and Eva~A. Alfoldi.
\newblock Progressive focusing and trustworthiness in qualitative research: The enabling role of computer-assisted qualitative data analysis software (caqdas).
\newblock {\em Management International Review}, 52(6):817--845, 2012.

\bibitem{trilling2018}
Damian Trilling and Jeroen~G. Jonkman.
\newblock Scaling up content analysis.
\newblock {\em Communication Methods and Measures}, 12(2--3):158--174, 2018.

\bibitem{vatsal2023}
Shubham Vatsal, Adam Meyers, and J.~Ortega.
\newblock Classification of us supreme court cases using bert-based techniques.
\newblock {\em Recent Advances in Natural Language Processing}, abs/2304.08649:1207--1215, 2023.

\bibitem{wachinger2024}
Jonas Wachinger, Kate Bärnighausen, Louis~N. Schäfer, Kerry Scott, and Shannon~A. McMahon.
\newblock Prompts, pearls, imperfections: Comparing chatgpt and a human researcher in qualitative data analysis.
\newblock {\em Qualitative Health Research}, 2024.

\bibitem{walgrave2019}
Stefaan Walgrave and Amber~E. Boydstun.
\newblock The comparative agendas project.
\newblock In {\em Comparative Policy Agendas: Theory, Tools, Data}, page~35. Oxford University Press, 2019.

\bibitem{zambrano2023}
Andres~Felipe Zambrano, Xiner Liu, Amanda Barany, Ryan~S. Baker, Juhan Kim, and Nidhi Nasiar.
\newblock From ncoder to chatgpt: From automated coding to refining human coding.
\newblock In {\em Communications in Computer and Information Science}, pages 470--485. Springer Nature Switzerland, Cham, 2023.

\end{thebibliography}
\end{document}